\documentclass[journal=cmatex,layout=twocolumn,manuscript=article]{achemso}

\makeatletter
\let\l@addto@macro\relax
\makeatother
\usepackage[fontsize=10pt]{scrextend}
\usepackage{charter}

\usepackage{type1cm}

\usepackage{microtype}
\usepackage[english]{babel}

\usepackage{etoolbox}
\usepackage{amsmath}
\usepackage{graphicx}
\usepackage{subfig}

\usepackage{geometry}
\usepackage{changepage}
\usepackage{ragged2e}
\usepackage{soul}
\usepackage{caption}
\renewcommand{\figurename}{\textbf{Figure}}
\renewcommand{\thefigure}{\textbf{\arabic{figure}}}

\usepackage{bm,bbm}
\usepackage[separate-uncertainty=true,range-phrase=--]{siunitx}
\usepackage{enumitem}
\usepackage{dcolumn}% Align table columns on decimal point
\newcolumntype{.}{D{.}{.}{-1}}
\usepackage{multirow,tabularx}
\AtBeginDocument{\usepackage{booktabs}}
\usepackage{xcolor}
\definecolor{darkblue}{rgb}{0.4, 0.4, 0.4}
\definecolor{citeblue}{rgb}{0.2305, 0.4102, 0.6172}
\usepackage{hyperref}
\hypersetup{colorlinks=true,citecolor=citeblue,urlcolor=citeblue,linkcolor=black}
\usepackage{memhfixc}
\usepackage{bibunits}
\defaultbibliographystyle{rsc}
\mciteErrorOnUnknownfalse

\newcolumntype{C}[1]{>{\centering\arraybackslash}p{#1}}
\usepackage{romanbar}
\newcommand{\RN}[1]{\textup{\uppercase\expandafter{\romannumeral#1}}}

\newcommand{\RefSI}[1]{{Supplementary \autoref{#1}}} %for referencing figures from the SI

\makeatletter
\newcommand\footnoteref[1]{\protected@xdef\@thefnmark{\ref{#1}}\@footnotemark}
\makeatother

\let\oldmaketitle\maketitle
\let\maketitle\relax

\author{\raggedright Adrian Ruckhofer}
\affiliation{\small\raggedright Institute of Experimental Physics, Graz University of Technology, Graz, Austria}
\author{Marco Sacchi}
\affiliation{Department of Chemistry, University of Surrey, Guildford GU2 7XH, United Kingdom}
\author{Anthony Payne}
\affiliation{Department of Chemistry, University of Surrey, Guildford GU2 7XH, United Kingdom}
\author{Andrew P. Jardine}
\affiliation{Cavendish Laboratory, J. J. Thompson Avenue, Cambridge CB3 0HE, United Kingdom}
\author{Wolfgang E. Ernst}
\affiliation{\small\raggedright Institute of Experimental Physics, Graz University of Technology, Graz, Austria}
\author{Nadav Avidor}
\affiliation{Cavendish Laboratory, J. J. Thompson Avenue, Cambridge CB3 0HE, United Kingdom}
\email{nadavavidor@gmail.com}
\author{Anton Tamt\"{o}gl}
\affiliation{\small\raggedright Institute of Experimental Physics, Graz University of Technology, Graz, Austria}
\email{tamtoegl@gmail.com}

\title{\raggedright\textrm{\Large \bfseries Evolution of ordered nanoporous phases during h-BN growth: Controlling the route from gas-phase precursor to 2D material by \emph{in-situ} monitoring.}}

\begin{document}

\begin{bibunit}
%%%%%%%%%%%%%%%%%%%%%%%%%%%%%%%%%%%%%%%%%%%%%%%%%%%%%%%%%%%%%
\twocolumn[
\begin{@twocolumnfalse}
\vspace*{-1cm}
\oldmaketitle
\vspace*{-0.4cm}
{\textcolor{darkblue}{\rule{\textwidth}{1pt}}}

\vspace*{0.2cm}
{\normalsize \textcolor{darkblue}{\textbf{ABSTRACT:}} Large-area single-crystal monolayers of two-dimensional (2D) materials such as graphene and hexagonal boron nitride (h-BN) can be grown by chemical vapour deposition (CVD). However, the high temperatures and fast timescales at which the conversion from a gas-phase precursor to the 2D material appear, make it extremely challenging to simultaneously follow the atomic arrangements. We utilise helium atom scattering to discover and control the growth of novel 2D h-BN nanoporous phases during the CVD process. We find that prior to the formation of h-BN from the gas-phase precursor, a metastable $(3\times3)$ structure is formed, and that excess deposition on the resulting 2D h-BN leads to the emergence of a $(3\times4)$ structure. We illustrate that these nanoporous structures are produced by partial dehydrogenation and polymerisation of the borazine precursor upon adsorption. These steps are largely unexplored during the synthesis of 2D materials and we unveil the rich phases during CVD growth. Our results provide significant foundations for 2D materials engineering in CVD, by adjusting or carefully controlling the growth conditions and thus exploiting these intermediate structures for the synthesis of covalent self-assembled 2D networks.\\[0cm]}
{\textcolor{darkblue}{\rule{\textwidth}{1pt}}}
\vspace*{0.4cm}
\end{@twocolumnfalse}
]

%#####################################################
\section*{Introduction}
\begin{figure}[htbp]
\centering
\includegraphics[width=0.49\textwidth]{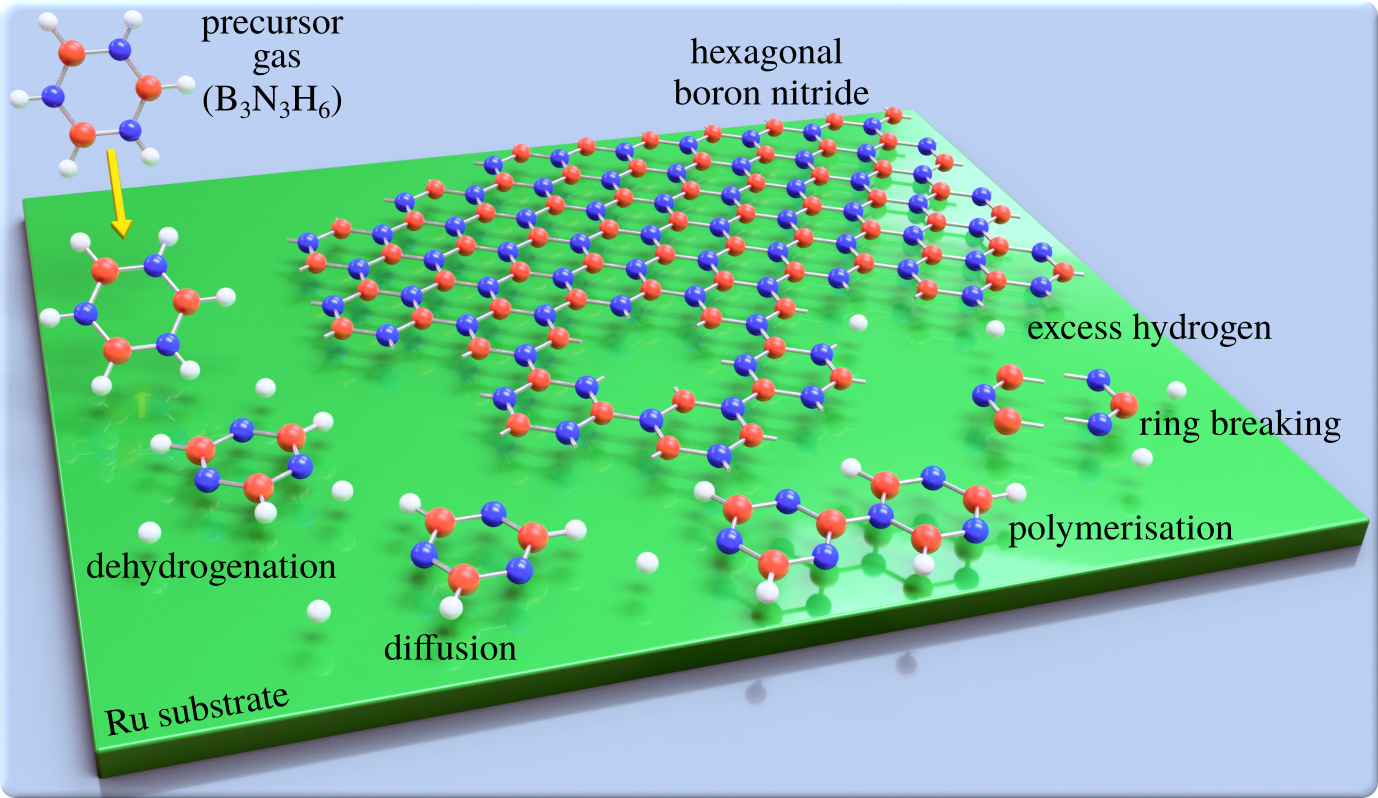}
\caption{Schematic illustrating the epitaxial growth of h-BN by chemical vapour deposition: A gaseous precursor (e.g. borazine, B$_3$N$_3$H$_6$) is brought into contact with a (hot) catalyst surface (Ru), triggering the chemical reactions such as breaking of the borazine rings and dehydrogenation, followed by the assembly of the epitaxial overlayer.}
\label{fig:TOC}
\end{figure}

%#####################################################
Two-dimensional (2D) materials such as graphene and hexagonal boron nitride (h-BN) offer technological promise$ $\cite{anichini2018,jiang2014}, e.g. with h-BN being considered to be the ``ideal'' dielectric for 2D based field-effect transistors\cite{ma2022}. However, their properties are highly dependent on the perfection of the 2D layers. For this reason, intense efforts have been devoted to study and improve the growth of defect-free 2D materials.\cite{zavabeti2020,zhang2017} A promising method of synthesising large-area 2D layers is chemical vapour deposition (CVD) and the CVD synthesis of atomically thin h-BN on metal substrates is described in several review articles.\cite{auwaerter2019,jana2018} The process, which is illustrated in \autoref{fig:TOC}, involves a gas-phase precursor deposited on a solid substrate at elevated temperatures. By diffusion and dehydrogenation or fragmentation of the precursor, the adsorbates are attached to growing clusters and eventually form the 2D layer. A complete dehydrogenation of the precursor requires overcoming multiple energy barriers. As a result, it might be expected that at intermediate temperatures, dehydrogenation would not be complete, which in-turn can result in metastable or intermediate structures. For the synthesis of bulk h-BN it is known that the process involves several steps of borazine-polymerisation.\cite{fazen1995,shi2010,bernard2014b,bernard2016} There are several routes, but even in the bulk the process has not been studied in great detail. Here, we follow a series of structural changes to identify  intermediate structures in 2D growth.
%#####################################################
%#####################################################
%Progress and potential
%The graphite-diamond phase transition is a central subject in physical science. Among the debates after many years of studies, one outstanding issue is the role of hexagonal diamond (HD), which was argued to be the preferable product according to the simulation but never reported in the compression experiments on graphite under high-pressure and high-temperature (HPHT) conditions. From a synergy between experiment and simulation, we investigated the atomistic mechanism of the graphite-diamond transition in HPHT conditions. Our study suggests that the growth of diamond has a preferred direction, which notably favors the formation of cubic diamond (CD). On the other hand, HD only appears as the twin structures of CD. We further investigated the possibility of harvesting the twin structures via microstructure engineering, which may have the potential to advance the fabrication process in the synthetic diamond industry.

While it is crucial to understand the growth process, mechanistic and kinetic studies are rare and mostly focus on the growth of nanocarbons.\cite{talyzin2011coronene,li2014carbon,chen2010contrasting,wang2022} Dehydrogenation and intermediate structures during CVD of 2D materials have been proposed,\cite{losurdo2011,zhang2014,lu2015,choi2017,qiu2018,habib2018rev,li2022} but to the best of our knowledge have not been studied experimentally. In general, kinetics and the thermochemistry of intermediate products, may lead to metastable structures. However, phase-diagrams due to partially dehydrogenated precursors have not been reported. Most studies report completed overlayer structures while the complexity and individual steps, as illustrated in \autoref{fig:TOC}, are often ignored. In particular, previous h-BN studies using real space methods\cite{goriachko2007,joshi2012,lu2013,steiner2019} concentrate on local order in completed h-BN structures, while reciprocal space studies\cite{mueller2009,martoccia2010,orlando2014} have provided information about long-range order.\cite{kelsall2021} 

In this paper we present a systematic analysis, at various temperatures beyond the ones reported for best growth conditions (\SIrange{1050}{1100}{\K}\cite{martoccia2010,lu2013,goriachko2007}) and at various dosing rates. By following h-BN growth \emph{in situ} using helium atom scattering (HAS) we demonstrate the existence of metastable structures during the formation of h-BN from borazine (B$_3$N$_3$H$_6$). HAS is a well-established technique for monitoring thin-film growth modes\cite{farias1998,braun1999,mcIntosh2014} and has been used to study the quality of CVD-grown 2D materials\cite{borca2010,tamtogl2015,anemone2016,gibson2016} and inter-layer interaction,\cite{taleb2018} yet investigations of intermediate structures have not been performed. In particular, we find that there is one precursor structure with a well-defined $(3\times3)$ periodicity, meaning a well-defined route for the polymerisation reaction which leads to h-BN. We further find that by dosing excess borazine, a $(3\times4)$ structure forms, which could be attributed to a partially polymerised second-layer on top of the formed h-BN.

Our experimental results are complemented by van der Waals (vdW) corrected density functional theory (DFT) calculations which confirm the nature of the system, helping us to determine which self-assembled structures are compatible with the experimental results.

%#####################################################
\section*{Results and Discussion}
%#####################################################
The adsorption of the precursor gas (borazine, B$_3$N$_3$H$_6$) on the Ru substrate has been investigated in several other studies using Auger electron spectroscopy, X-ray photoelectron spectroscopy, electron energy loss spectroscopy and low energy electron diffraction.\cite{paffett1990,orlando2012,farkas2015} There is general consensus in the literature, that borazine only adsorbs molecularly at low (<\SI{140}{\K}) temperatures\cite{he1990,orlando2012,farkas2015} with dehydrogenation setting in at temperatures of 150-\SI{250}{\K}, depending on the substrate.\cite{orlando2012,farkas2015,haug2020}
Starting from about \SI{600}{\K}, again depending on the metallic substrate, the B-N ring is reported to break down into its atomic constituents.\cite{he1990,haug2020} According to Paffet \textit{et al.} \SI{1000}{\K} are necessary for h-BN formation on Ru(0001)\cite{paffett1990,orlando2012} while hydrogen desorption occurs over a wide temperature range\cite{farkas2015} and may even intercalate in the h-BN layer.\cite{spaeth2017,kim2020}

Helium diffraction allows \emph{in situ} measurements even at growth temperature, and is known for its unique sensitivity to adsorbates, including hydrogen atoms.\cite{lau2018structural,avidor2011highly,kraus2016,bahn2016,avidor2016,lin2018,tamtogl2020,tamtogl2021a,tamtogl2021} Furthermore, unlike other established techniques,\cite{lin2015} HAS is completely inert and does not modify the process under investigation.\cite{benedekTB} While the specular reflection gives an estimate of adsorbate coverage on the clean surface, the angular distribution provides insight in the time evolution of periodic structures being formed on the surface.\cite{tamtogl2015,kelsall2021} In the present work, CVD growth was performed at a set crystal temperature while monitoring the surface using repeated one-dimensional angular diffraction scans, where we observe the emergence and disappearance of additional superstructures followed by the formation of h-BN.

All experimental data was obtained with the Cambridge spin-echo apparatus which uses a nearly monochromatic atomic beam of $^3$He,  scattered off the sample in a total scattering angle of \SI{44.4}{\degree} and with an incident energy of \SI{8}{\meV} (see \nameref{sec:SIExp} in the Supporting Information). The parallel momentum transfer $\Delta K$ is given by $\Delta K = \lvert \Delta \mathbf{K} \rvert = \lvert \mathbf{K}_f - \mathbf{K}_i \rvert = \lvert \mathbf{k}_i \rvert \left( \sin \vartheta_f  - \sin \vartheta_i  \right)$, with $\mathbf{k}_i$ being the incident wavevector and $\vartheta_i$ and $\vartheta_f$ the incident and final angles with respect to the surface normal, respectively.\cite{jardine2009,Alexandrowicz2007,holst2021} Compared to techniques such as scanning tunnelling microscopy (STM), HAS averages over larger surface areas, typically $\approx\SI{3}{\milli\meter\squared}$. Therefore, the advantage of HAS is to give precise information about any long-range periodicity of surface structures. For ease of comparison, the diffraction scans for the different structures are plotted as a function of the parallel momentum transfer, $\Delta K$, relative to the $G_{01}$ peak of the Ru(0001) substrate, $|\Delta K/G_{01}| $. By converting the abscissa in this way, the position of the observed diffraction peaks directly reflects their periodicity with respect to the substrate lattice spacing.
%#####################################################
\begin{figure}[h]
\centering
\includegraphics[width=0.45\textwidth]{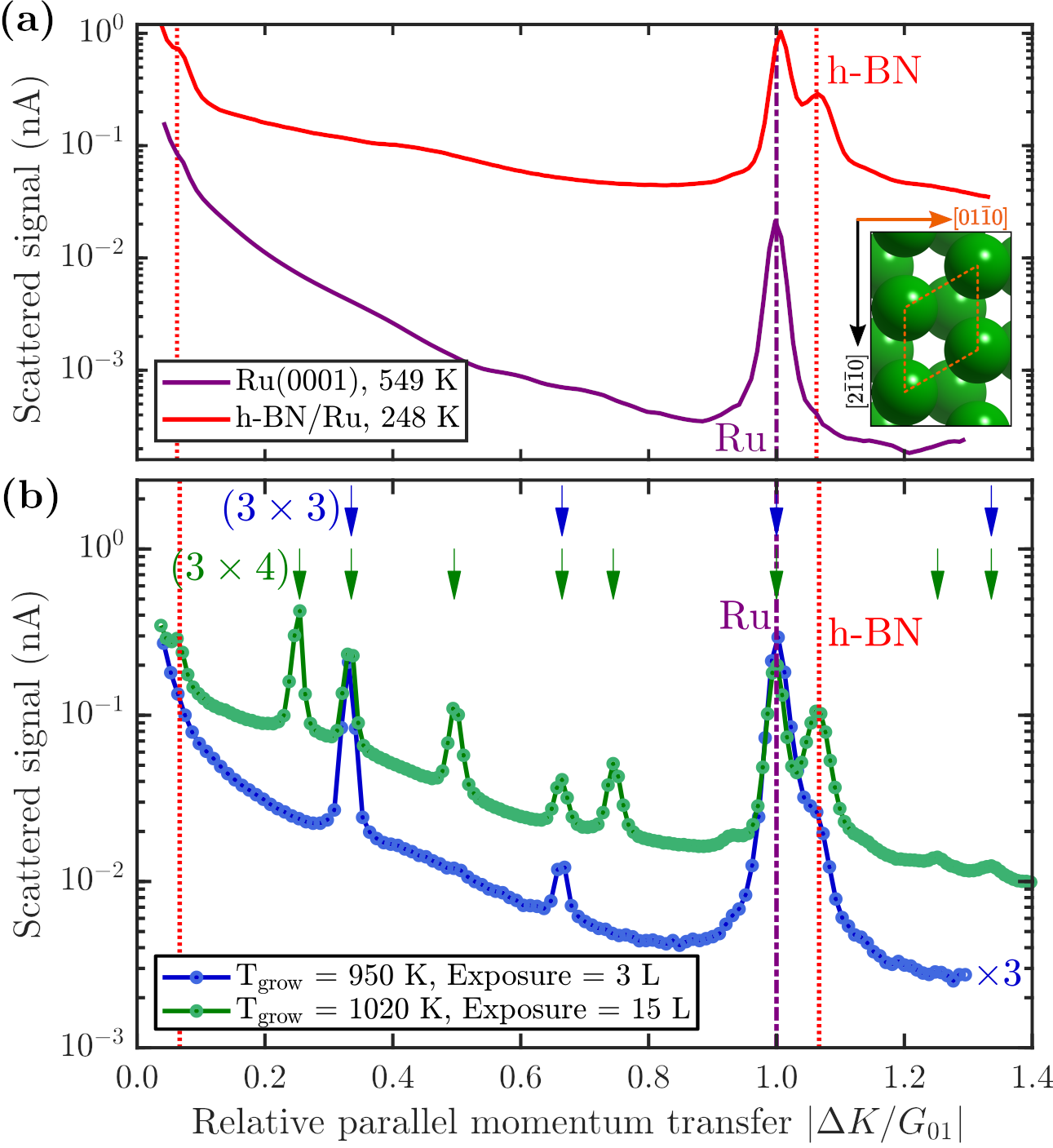}
\caption{(a) Comparison of the angular diffraction scans for clean Ru(0001) in purple and the completed h-BN overlayer on Ru in red. The $\langle 01\bar{1}0\rangle$ scanning direction ($\overline{\Gamma\mathrm{M}}$) is shown in the inset. The purple dash-dotted line indicates the position of the first order Ru diffraction peak and the red dotted lines the h-BN Moir{\'e} peaks. (b) Diffraction scans during borazine exposure reveal additional superstructures, depending on the growth temperature and total exposure. The position (shown by the arrows at the top) and spacing of the additional peaks reveal a $(3\times3)$ / $(3\times4)$ superstructure plotted in blue / green. Low exposure at lower temperatures reveals a $(3\times3)$ structure (blue curve, grown at \SI{950}{\K}), while at higher exposures 
and higher temperatures an intermediate $(3\times4)$ pattern emerges (green curve, grown at \SI{1020}{\K}). To improve the signal to noise ratio, the sample was subsequently cooled down for the duration of both scans and the blue curve was scaled by a factor of 3 to facilitate comparison.}
\label{fig:superstruct}
\end{figure}
%#####################################################

\subsection*{A precursor structure to h-BN growth}\label{sec:ResA}
First, we describe how borazine exposure at low temperature $(T < \SI{880}{\K})$ reveals a precursor structure on Ru(0001), which by further annealing at $T = \SI{880}{\K}$ can be converted to h-BN.

The purple line in \autoref{fig:superstruct}(a) shows a diffraction scan of the clean Ru(0001) substrate. Exposing Ru to 7\,\mbox{Langmuir} (L) of borazine at a surface temperature of \SI{600}{\K}, results in decreased diffraction intensities and helium reflectivity. Moreover, the lack of any additional diffraction peaks is typical of a disordered structure.\cite{tamtogl2021} The behaviour is consistent with earlier studies showing that the B-N ring starts to break down into its constituents only above about \SI{600}{\K}.\cite{he1990,haug2020,paffett1990}

Upon increasing the temperature to \SI{750}{\K} while maintaining borazine overpressure, additional peaks start to appear between the specular and first order Ru diffraction peaks. \autoref{fig:superstruct}(b) (blue curve) shows the characteristic diffraction pattern that emerges. Equidistant peaks at $|\Delta K/G_{01}| =  0.33$ and $0.66$ indicate a $(3\times3)$ periodic structure on the surface, which we label BN$_{\RN{1}}$. If dosing is performed at even higher temperatures ($T \geq \SI{880}{\K}$), in addition to the observed BN$_{\RN{1}}$ structure, a shoulder appears to the right-hand side of the first order Ru diffraction peak, indicating the formation of a h-BN structure on the surface (vertical red dotted line in \autoref{fig:superstruct}). The peak, which occurs at $|\Delta K/G_{01}| =  1.08$, is a result of the commensurate Moir{\'e} pattern on Ru\cite{martoccia2010} (see also \hyperref[sec:SISuper]{h-BN periodicity} in the SI).

\begin{figure}[htbp]
\centering
\includegraphics[width=0.48\textwidth]{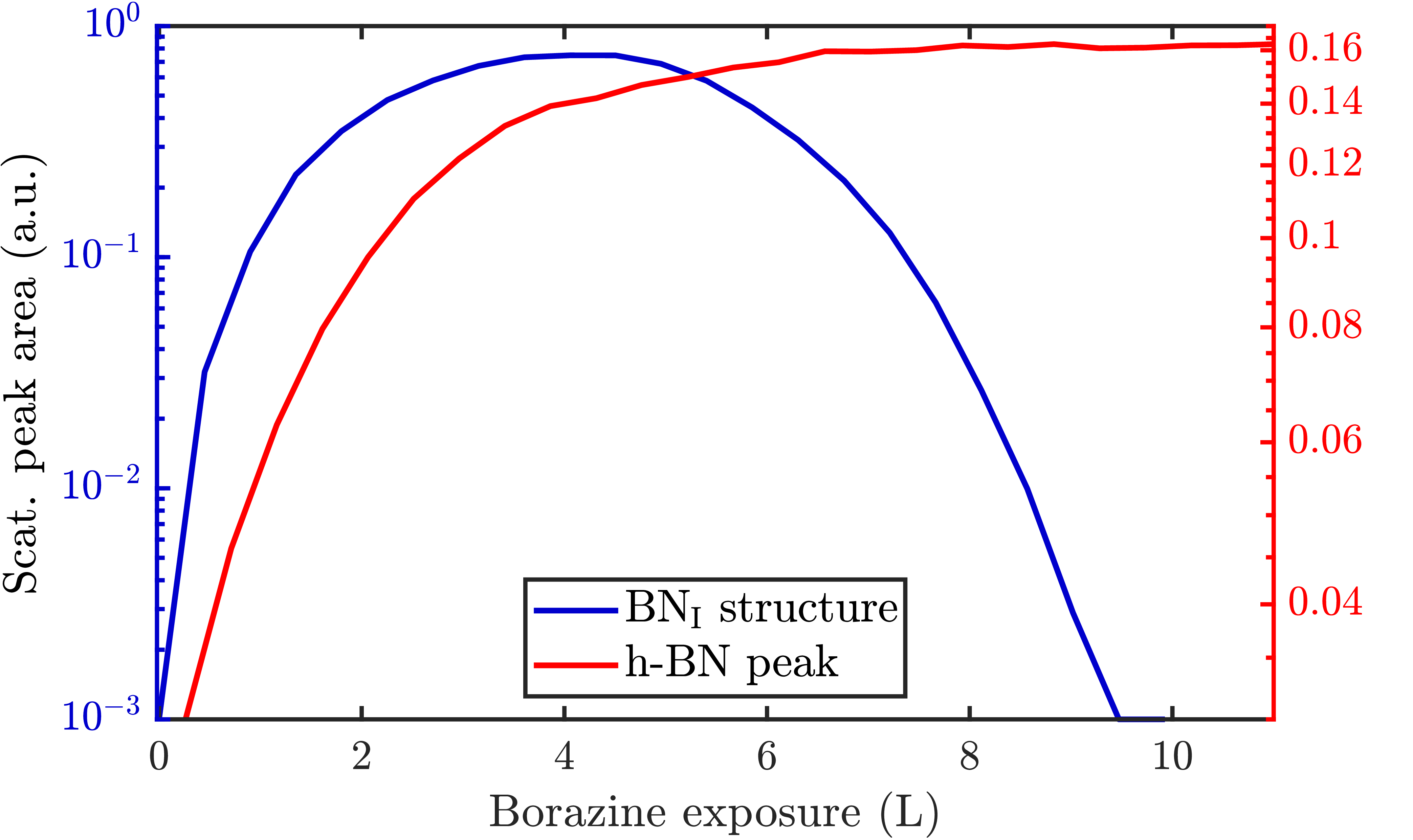}
\caption{\emph{In situ} monitoring of the integrated peak intensities reveals that the BN$_{\RN{1}}$ structure acts as precursor to the h-BN overlayer. The characteristic diffraction peaks for the BN$_{\RN{1}}$ structure and the h-BN peak are plotted versus borazine exposure at a substrate temperature of \SI{880}{\K}. The BN$_{\RN{1}}$ structure increases prior to the h-BN intensity and has already disappeared when the h-BN intensity exhibits its maximum.}
\label{fig:phasediag_3x3}
\end{figure}

Since a $(3\times3)$ superstructure composed of intact borazine molecules shows only weak binding to the substrate, the observed BN$_{\RN{1}}$ structure, as shown later in our DFT calculations, must be composed of partly dehydrogenated borazine molecules, in line with the reported low experimental dehydrogenation temperature on other substrates.\cite{haug2020} \autoref{fig:phasediag_3x3} illustrates \emph{in situ} monitoring of the integrated peak intensities, which demonstrates that the BN$_{\RN{1}}$ structure precedes the growth of h-BN. The exposure dependent intensities are obtained from repeated angular diffraction scans. Immediately after dosing begins, the $(3\times3)$ peaks start to rise rapidly (blue line), while only after a short delay the h-BN diffraction peak increases (red line), although less quickly than the BN$_{\RN{1}}$ structure. The h-BN peak intensity reaches its maximum at the same point where the BN$_{\RN{1}}$ structure disappears. We conclude that the BN$_{\RN{1}}$ structure is converted into h-BN and acts as a precursor structure to the complete h-BN overlayer. Since the intensity of the BN$_{\RN{1}}$ structure drops to almost zero at $\approx$\SI{9}{\L}, it indicates that virtually all $(3\times3)$ domains are converted to h-BN.
%#####################################################
%#####################################################

%We suggest that the BN$_{\RN{1}}$ structure is formed from partially dehydrogenated borazine molecules, since
The temperature region where the BN$_{\RN{1}}$ structure evolves is in excellent agreement with the desorption temperature of hydrogen reported by Paffett \emph{et al.},\cite{paffett1990} which helps confirm the dehydrogenation process. Further, as mentioned earlier, bulk h-BN is known to form by a sequence of dehydrogenation processes, in which borazine polymerises to polyborazylene, which is then cross-linked in one or more steps.\cite{shi2010,bernard2014b} Our results suggest that a similar process happens at the ruthenium surface, but that in the 2D case, there is one clear intermediate step, i.e. the BN$_{\RN{1}}$ structure, before the formation of h-BN. There are several possible real space structures, which are in line with the observed periodicity and composition\cite{li2022}. We concentrate on providing experimental proof that such intermediate / precursor structures with long-range order exist, while the exact chemical composition of these structures is better probed with X-ray photoemission spectroscopy (XPS) studies\cite{orlando2012}, beyond the mentioned structural analysis. Moreover, the precursor gas imposes a B/N ratio of 1:1 which we do not expect to change and we have therefore focussed on supporting our structural characterisation through detailed DFT modelling of appropriate candidate structures.
%There are several possible real space structures, which are in line with the observed periodicity and composition. The precursor imposes a B/N ratio of 1:1 which we do not expect to change. To chemically characterise the structure, X-ray photoemission spectroscopy (XPS) would typically be used for chemical characterisation of CVD-grown h-BN.\cite{orlando2012} However, it can be difficult to obtain chemical information regarding the hydrogenation state from XPS.  We have therefore focussed on supporting our structural characterisation through detailed DFT modelling of appropriate candidate structures.

\subsection*{DFT structural modelling}\label{sec:DFT}
We understand the BN$_{\RN{1}}$ structure to consist of partially dehydrogenated, polymerised borazine analogous to the synthesis of bulk h-BN. Thus it is apparent that the 2D growth occurs in a step-by-step process and that not all hydrogen atoms are expected to be removed at the same time, due to the different bonding strengths to N and B atoms, as well as the bond formation between the N and the Ru atoms. Based on this attribution, we have made a vdW-corrected DFT investigation into the energetics of the BN$_{\RN{1}}$ structure, from adsorption of the precursor gas to the complete h-BN overlayer.\\
We start by considering a single borazine molecule in a $(3\times3)$ supercell (see \nameref{sec:DFTMethods} in the Supporting Information), and move on to partially dehydrogenated borazine polymers. From the adsorption energies of isolated borazine molecules we observe that the bonding becomes much stronger with dehydrogenation, but the calculations cannot provide a definitive answer in terms of the dehydrogenation sequence (dehydrogenation of the B atoms is slightly more favourable than of the N atoms by $\approx$\SI{15}{meV}). However, as shown later for two borazine molecules per supercell, the N atoms can dehydrogenate more easily than the B atoms and the candidate structures for our observations can be clearly distinguished in terms of the adsorption energies. 

\begin{figure*}[htp]
\centering
\includegraphics[width=0.98\textwidth]{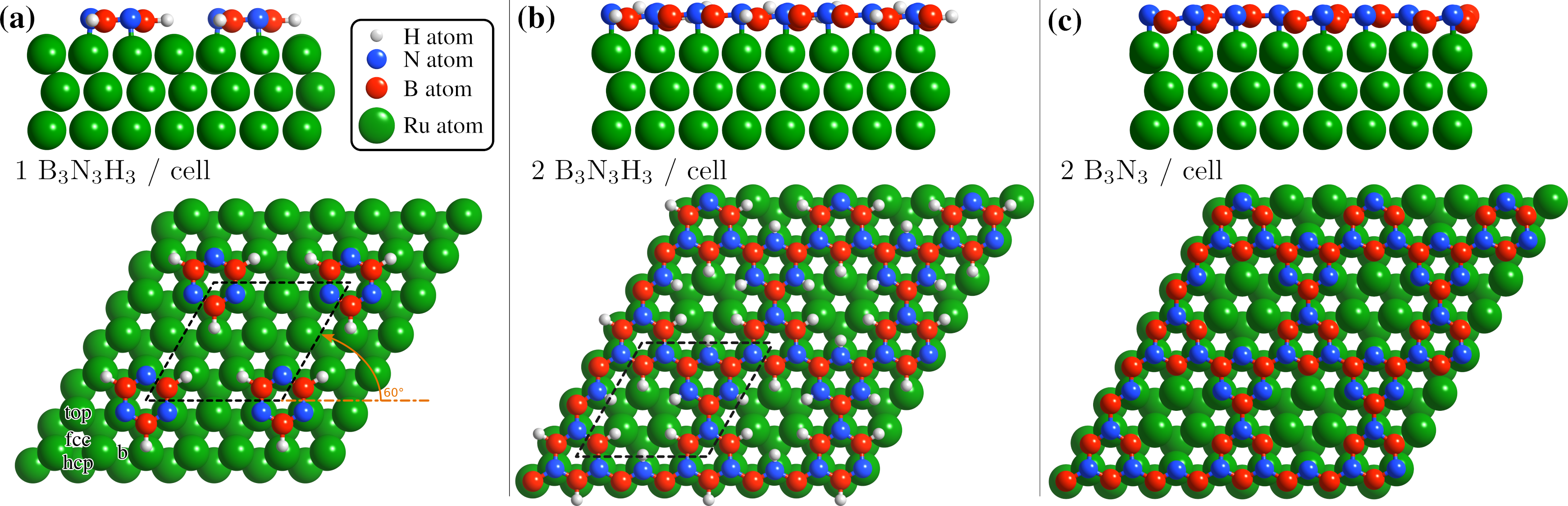}
\caption[]{Side and top view of the energetically most favourable configurations on the Ru(0001) surface, from one to two borazine molecules per supercell. (a) For one partially dehydrogenated borazine molecule (B$_3$N$_3$H$_3$) adsorption occurs on the fcc site, forming a $(3\times3)$ structure. (b) shows the structure for two partially dehydrogenated B$_3$N$_3$H$_3$ molecules per supercell, where bound B-N rings form a nanostructured network and hydrogen atoms remain adsorbed on the Ru lattice inside the nano\-pores. In (c) the optimised structure for two fully dehydrogenated borazine molecules is shown, leading to the same nanopore structure with a $(3\times3)$ periodicity.}
\label{fig:BorAds}
\end{figure*}

%#################################################
\begin{table}[htbp]
\centering
\caption{DFT calculations for the adsorption structures of the borazine precursor on Ru(0001), based on one molecule per $(3\times3)$ supercell. The results are shown for an intact (B$_3$N$_3$H$_6$) and partially dehydrogenated (B$_3$N$_3$H$_3$) adsorbate, considering various initial adsorption sites and a rotation of 60$^{\circ}$ (see  \autoref{fig:BorAds}(a)). The adsorption energies E$_\mathrm{ads}$ are given for the final optimised adsorption sites and $\Delta$E is the difference with respect to the minimum energy configuration of the system with the same dehydrogenation state.}

\resizebox{.47\textwidth}{!}{
\begin{tabular}{ c | c | c || c | c | c }
	\toprule
	\multicolumn{3}{c||}{\textbf{B$_3$N$_3$H$_6$} } & \multicolumn{3}{c}{\textbf{B$_3$N$_3$H$_3$}} \\[1mm]
	\midrule
	\rule{0mm}{5mm}
	Site & E$_\mathrm{ads}$ (eV)  & $\Delta$E (eV)   & Site & E$_\mathrm{ads}$ (eV)  & $\Delta$E (eV)\\ [1mm]
	\midrule
	\rule{0mm}{5mm}
	 fcc & -4.08 & 0.00 & fcc & -8.95  & 0.00 \\
	 top  & -1.21  & 2.86 &   top & -6.60  & 2.35  \\
  b$\rightarrow$fcc  & -4.08  & 0.00 &   b$\rightarrow$fcc & -8.95  & 0.00 \\
	  hcp  & -1.31 & 2.77 &  hcp  & -5.88  & 3.07 \\[1mm]
	\bottomrule
	\end{tabular}}
\label{tab:1BZ_ads}	
\end{table}
%################################################# 
In \autoref{tab:1BZ_ads} we compare the binding energies from vdW-corrected DFT for an intact (B$_3$N$_3$H$_6$) and a partially dehydrogenated (B$_3$N$_3$H$_3$) borazine molecule with one molecule per $(3\times3)$ supercell, confirming a much stronger bonding of B$_3$N$_3$H$_3$. We consider various initial adsorption sites (\autoref{fig:BorAds}(a)) with respect to the C$_3$ rotational axis through the centre of the molecule and a rotation of 60$^\circ$. Adsorption occurs in a flat face-to-face configuration, while bonding of the same adsorbates with a rotation of 0$^\circ$ is slightly weaker -- the results are shown in the Supplementary Information (see \nameref{sec:SIDFT}).

In \autoref{tab:1BZ_ads} the energy differences $\Delta$E are given with respect to the minimum energy of the same dehydrogenation state in addition to the respective adsorption energies E$_\mathrm{ads}$. For both stoichiometric configurations the most favourable position is the fcc site and if the borazine molecule is initially placed on a bridge site it undergoes a transition to this position. The fcc configuration for partially dehydrogenated borazine yields an adsorption energy of E$_\mathrm{ads}$=\SI{-8.95}{\eV} and is shown in \autoref{fig:BorAds}(a) with the $(3\times3)$ supercell highlighted by the black dashed rhombus. The results for the intact borazine molecule (B$_3$N$_3$H$_6$) are very similar with respect to the adsorption site, however, we obtain significantly weaker bonding strengths compared to the dehydrogenated molecule.

Based on bulk h-BN studies we conclude that it is more likely that polymerised networks are formed.\cite{shi2010,bernard2014b} Starting from the minimum energy configuration of a single borazine molecule on the fcc site we continue by adding a second borazine molecule in the supercell. By considering various initial rotations of the additional molecule the energetically most favourable configurations were then identified. In contrast to the case of an isolated borazine molecule, the dehydrogenation sequence becomes clearly discernible in terms of the adsorption energies, with two borazine molecules. The N atoms in the ring adsorb on top of the Ru atoms and the borazine molecules lose all  hydrogen atoms associated with the N atoms upon bond formation, in line with experimental results of the completed h-BN overlayer where inter-layer bonding is facilitated via the N atoms.\cite{auwaerter2019,goriachko2007} Such a scenario is, however, different to bulk h-BN growth where ruthenium is not present and thus interaction with the substrate may give rise to an even faster loss of hydrogen compared to bulk studies.

The calculations for two intact borazine molecules per supercell (2B$_3$N$_3$H$_6$, not shown) yield weak binding, since the H atoms start to overlap resulting in a tilt of the complete molecules with respect to the surface. Moreover, due to desorption of hydrogen atoms from the borazine at low temperatures it is unlikely that intact borazine will remain and so intact molecules will not be considered further.\cite{paffett1990} 

Therefore, we concentrate on partially and fully dehydrogenated borazine molecules. \autoref{fig:BorAds}(b,c) shows the final optimised structure for 2B$_3$N$_3$H$_3$ per supercell, illustrating that individual borazine molecules form bonds to each other. The bound B-N rings build up a nanostructured network with nanopores, i.e. where in between the B-N rings vacancies/pores of the Ru substrate are left behind. The high binding energy of the structure in \autoref{fig:BorAds}(b) with \SI{-6.28}{\eV} compared to \SI{-6.74}{\eV} for the complete h-BN/Ru, may therefore explain the stability of the BN$_{\RN{1}}$ structure at temperatures $\approx$\SI{750}{\K} as observed in the experiments. 
%#####################################################
%#####################################################

The structure in \autoref{fig:BorAds}(b) acts as an intermediate prior to complete dehydrogenation which is expected at elevated temperatures. The calculations show that first the hydrogen atoms detach from the nitrogen and bind to the Ru substrate on the hcp sites, inside the nanopores. The excess hydrogen adatoms inside the nanopores are likely to desorb relatively quickly at the temperature of the experiment.\cite{Feulner1985} Therefore, \autoref{fig:BorAds}(c) shows the optimised structure, starting with two fully dehydrogenated borazine molecules per supercell, leading again to the formation of nanopores. Such an open structure could easily act as a precursor to the complete h-BN overlayer, since each pore only has to be ``filled'' with an additional dehydrogenated borazine molecule. Finally, the addition of further borazine molecules in the calculations, i.e. three per supercell, essentially leads to the formation of h-BN which gives rise to the strongest binding energy in the calculations. The route from the precursor BN$_{\RN{1}}$ structure to the final h-BN overlayer, with several intermediate steps, is illustrated in \RefSI{fig:hBNRoute}.

In addition to providing us with real-space structures of the observed BN$_{\RN{1}}$ precursor, there are several points which we note from the vdW DFT calculations:
Dehydrogenation of borazine always gives rise to a stronger bonding to the substrate and the results show that the thermodynamically most stable configuration for three adsorbed borazine molecules is h-BN (\RefSI{fig:DFT1}(b)).
We also see from the side views in \autoref{fig:BorAds} that there occurs always some buckling ($0.21-\SI{0.35}{\angstrom}$) and the adlayer is never perfectly flat. The results show that by carefully controlling the substrate temperature and thus the amount of excess hydrogen in future experiments, several BN nano-structures could be synthesized as shown for two cases in \autoref{fig:BorAds}(b,c). Moreover, careful changes of the starting conditions in the DFT calculations may even yield a ``local'' minimum energy configuration as in \RefSI{fig:DFT1}(c). Thus the system may be an ideal playground for the growth of different nano-structures and further metastable networks beside the ones reported in this work. %may well exist.

\subsection*{Additional structures accompanying the h-BN growth}\label{sec:ResB}
So far, we have described the formation of a BN$_{\RN{1}}$ structure at $T \geq \SI{750}{\K}$, which is converted to h-BN at $T \geq \SI{880}{\K}$. However, upon complete conversion of the BN$_{\RN{1}}$ structure to h-BN, exposing the surface to excess borazine results in the emergence of an additional structure with a $(3\times4)$ periodicity, which we label BN$_{\RN{2}}$. The green line in \autoref{fig:superstruct}(b) illustrates the corresponding diffraction pattern with the h-BN Moir{\'e} diffraction peak being still present next to the first order Ru peak. As shown in a two-dimensional diffraction scan in \RefSI{fig:2dscan}, the $(3\times4)$ peaks are not a subset of the h-BN Moir{\'e} pattern. In addition, a smaller peak to the left of the first order Ru peak becomes visible which can be attributed to a substrate reconstruction peak\cite{martoccia2010} due to the h-BN growth.

\begin{figure}[htb]
\centering
\includegraphics[width=0.47\textwidth]{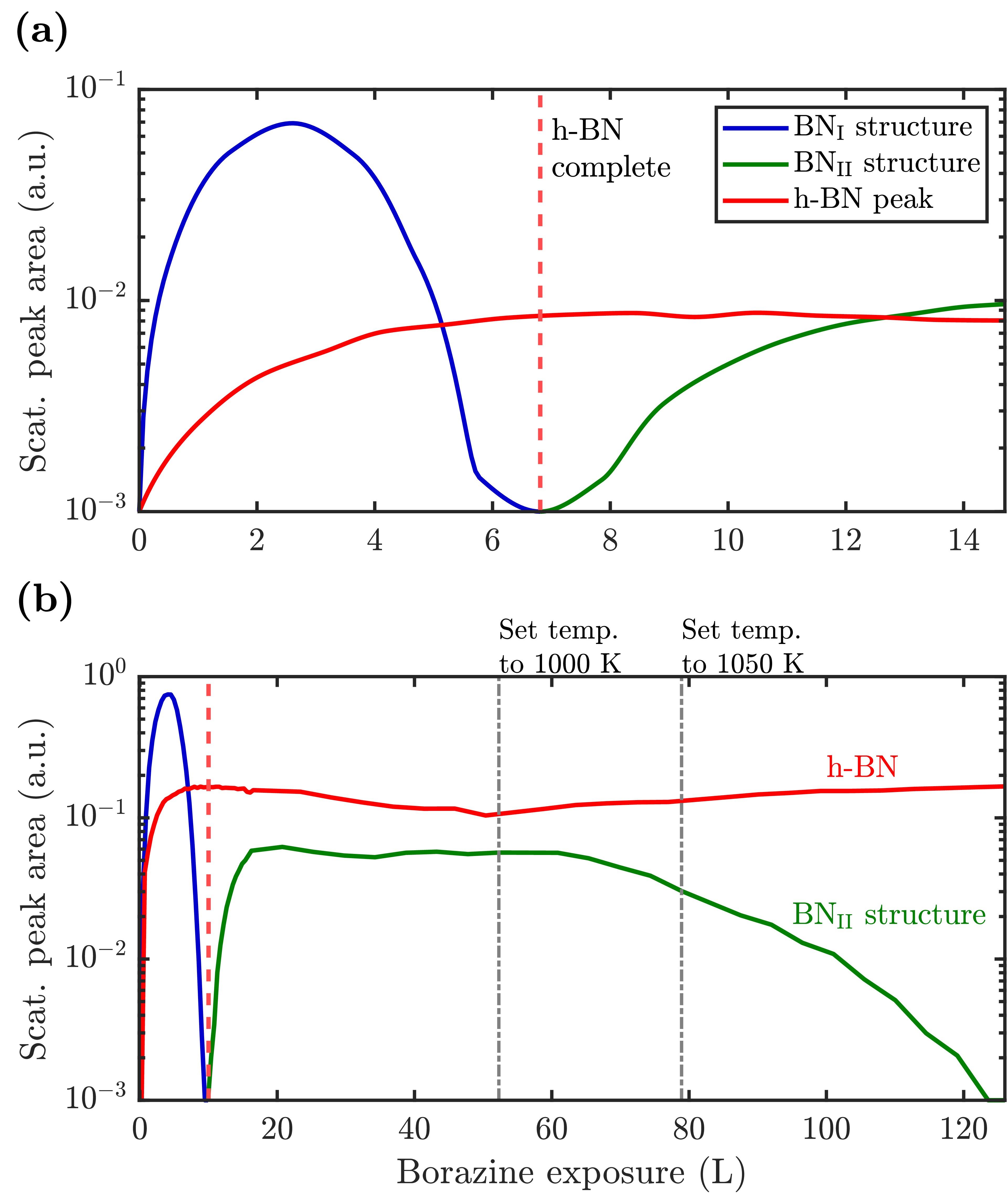}
\caption{Peak areas of the characteristic diffraction peaks representing the different structures versus borazine exposure. After the BN$_{\RN{1}}$ structure has disappeared the h-BN peak saturates giving rise to a conversion and further exposure leads to the rise of the BN$_{\RN{2}}$ structure. Due to a higher substrate temperature of \SI{915}{\K} in (a), the BN$_{\RN{1}}$ structure disappears already after an exposure of $\approx$\SI{7}{\L} compared to \SI{10}{\L} at \SI{880}{\K} in (b), thus indicating a kinetically driven conversion. Dosing in (b) is then further continued with subsequent changes of the surface temperature as stated above the diagram. After long enough exposure the BN$_{\RN{2}}$ structure disappears leaving a strong h-BN intensity behind.}
\label{fig:phasediag}
\end{figure}

To monitor the growth of the BN$_{\RN{2}}$ structure we use a smaller borazine overpressure while holding the sample temperature at \SI{915}{\K}. \autoref{fig:phasediag}(a) shows the evolution of the BN$_{\RN{2}}$ and the BN$_{\RN{1}}$ structure as blue and green curves, respectively. Here, the red line is again the integrated peak intensity of the h-BN diffraction peak. Immediately after exposing the surface to borazine, the BN$_{\RN{1}}$ structure increases together with the h-BN peak. Further exposure leads to a decay of the BN$_{\RN{1}}$ structure, while the h-BN feature still rises, indicating the growth of h-BN islands. At \SI{7}{\L} the h-BN diffraction peak saturates, while at the same time the BN$_{\RN{1}}$ structure disappears. At this stage the h-BN overlayer is complete and after further dosing of borazine, the BN$_{\RN{2}}$ structure starts to emerge. As discussed later this may be interpreted as a second layer being formed on top of h-BN. 

The measurement was repeated at an even lower dosing pressure, while holding the sample at the lower temperature of \SI{880}{\K}. In \autoref{fig:phasediag}(b) the same behaviour is reproduced, yielding a h-BN layer with two additional structures, except that the emergence is delayed to longer/higher exposures, thus indicating a kinetically driven conversion.

With continuing borazine exposure to \SI{20}{\L} in \autoref{fig:phasediag}(b), the BN$_{\RN{2}}$ structure reaches its maximum with no further changes in the scattered intensity. Together with the rise of the BN$_{\RN{2}}$ structure the h-BN peak intensity slowly starts to decay, likely due to diffuse scattering from additional adsorbates at the surface or from domain walls of the BN$_{\RN{2}}$ structure. Increasing the surface temperature to \SI{1000}{\K} gives rise to a decay of the BN$_{\RN{2}}$ structure while the h-BN peak intensity starts to recover to its original value. Further temperature increase accelerates this process giving rise to a faster transition/conversion until the intermediate peaks disappear, leaving behind only the h-BN layer. Such a behaviour illustrates that ultimately h-BN is the most stable structure. Even though the borazine overpressure was still present, no additional peaks formed and the h-BN overlayer is the only remaining structure at the surface.

%#####################################################
%#####################################################
From \autoref{fig:phasediag}(b) it becomes evident, that in contrast to the BN$_{\RN{1}}$ structure, the BN$_{\RN{2}}$ structure is much more stable at higher temperatures since the $(3\times4)$ diffraction peaks are observed up to \SI{1000}{\K}. Further increase of the temperature to $\approx\SI{1200}{\K}$ gives rise to the surface migration of bulk-dissolved carbon, leading to the formation of graphene (see \nameref{sec:SIDiff1}) and thus eventually destroys the h-BN overlayer. The latter may open up the possibility to study the growth of h-BN/graphene heterostructures\cite{sutter2012,yin2016,zhang2015,kutana2015} but is beyond the scope of the current study.

From our experiments it is likely that the BN$_{\RN{2}}$ structure is a second chemisorbed layer on top of already grown h-BN. Earlier works on an Ir(111) substrate showed the evolution of additional compact reconstructed regions with a $(6\times2)$ superstructure, which were attributed to reconstructed boron areas.\cite{petrovic2017} On the other hand, CVD growth on polycrystalline Cu provided evidence for boron dissolution into the bulk together with multilayer h-BN formation via intercalation.\cite{kidambi2014} However, both systems and studies are significantly different from our approach. E.g., the different behaviour in the first study could be due to changes of both the lattice constant and the h-BN-substrate bonding between Ru and Ir. Moreover, in light of the recent observation of h-BN multilayer growth\cite{ma2022}, a second chemisorbed layer is much more plausible.

The BN$_{\RN{2}}$ structure, as a second chemisorbed layer, consists of partly or completely dehydrogenated borazine molecules with a desorption temperature slightly above our performed measurements, since we see an adsorption/desorption equilibrium at temperatures $<\approx\SI{1000}{\K}$ with an ultimate desorption at temperatures above this value. The existence of such a structure might be a precursor to multilayer synthesis if the original h-BN layer is of poor quality providing a high density of growth nuclei and thus explaining the reports of multilayer growth.\cite{song2010,guo2012,kidambi2014,gilbert2019,ma2022}

In a set of additional DFT calculations summarised in \RefSI{tab:DFT_hBN}, we considered also the possibility of borazine adsorption on top of h-BN/Ru as well as the formation of bi-layer h-BN.\cite{song2010,guo2012,kidambi2014,gilbert2019} However, we can rule out the latter according to our deposition measurements, since we do not detect oscillations of the BN$_{\RN{2}}$ structure or observations of any other periodicity, that would be indicative of multilayer h-BN growth. In line with the multi-stage process of h-BN bulk formation it is more likely that the BN$_{\RN{2}}$ structure consists of adsorbed molecules or polymerised borazine structures - with a weaker bonding compared to the first h-BN layer and therefore more likely to desorb. Further unlikely scenarios are discussed in \nameref{sec:SIDiscus}.

\subsection*{h-BN growth diagram on Ru(0001)} 
\begin{figure}[htbp]
\centering
\includegraphics[width=0.47\textwidth]{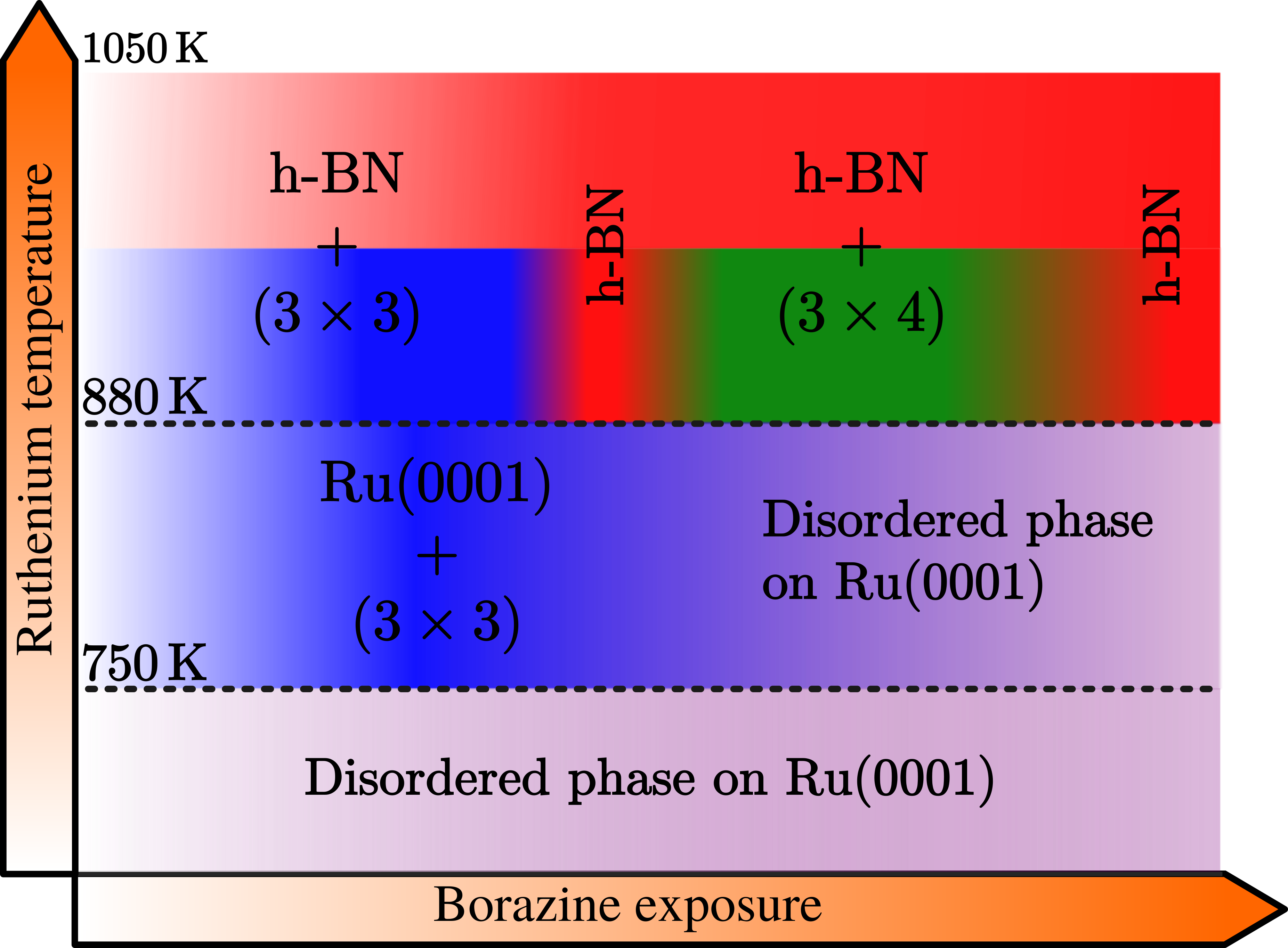}
\caption{Schematic growth diagram of the different BN structures observed on a Ru surface within various temperature ranges and with increasing borazine exposure. At temperatures below \SI{750}{\K} no periodic structure is formed, while above, a $(3\times3)$ pattern is observed, disappearing again at higher exposures. Above \SI{880}{\K} the $(3\times3)$ structure precedes the growth of h-BN, with the temperature being necessary for the formation of h-BN. With increasing borazine exposure, the latter is followed by a structure with $(3\times4)$ periodicity with respect to the Ru lattice, leading again to h-BN after long enough exposure. (See text for a precise explanation of the structures.)}
\label{fig:schem_phasediag}
\end{figure}

The combination of measurements and  DFT calculations allows us to conclude that the whole system passes through various structural phases:
{\small
\begin{equation*}
\mbox{Ru}+\mbox{BZ} \rightarrow \mbox{BN}_{\RN{1}}\,+\, \mbox{h-BN} \rightarrow \mbox{h-BN} \rightarrow \mbox{BN}_{\RN{2}}\,+\,\mbox{h-BN} \rightarrow \mbox{h-BN} \, ,
\end{equation*}}
with the outcome depending strongly on substrate temperature, borazine exposure and the point where one stops. In particular, the surface temperature strongly influences the kinetics and thus the duration and appearance of the additional superstructures. Combining the experimental results we derive a growth diagram as shown in \autoref{fig:schem_phasediag}, which describes the phenomenology of various structures arising during the CVD growth of h-BN on Ru(0001). 
%###########################################################
%###########################################################

Below \SI{750}{\K} no periodic overlayer structure on the Ru(0001) surface is found. Between $750$ and \SI{880}{\K} the BN$_{\RN{1}}$ structure forms on the surface which upon further borazine exposure vanishes and leaves a disordered phase behind. The minimum temperature to form a h-BN overlayer on the surface was determined to be \SI{880}{\K}. Above this temperature we observe additional structures, starting with a $(3\times3)$ structure (BN$_{\RN{1}}$) followed by a $(3\times4)$ periodic diffraction pattern (BN$_{\RN{2}}$). These structures always appear in addition to the h-BN layer and ultimately vanish, leaving a complete h-BN overlayer behind (see \autoref{fig:superstruct}(a) for a diffraction scan of a complete h-BN overlayer without any additional structures). We have thus identified two kinetic barriers which need to be overcome in order to form ordered structures on the Ru substrate: A temperature of $750\,$K is necessary for the precursor structure to result, while at $880\,$K the h-BN formation sets in. As mentioned above, the BN$_{\RN{1}}$ precursor structure is always present, however, with increasing temperature its transformation into h-BN becomes faster.

As further illustrated in \hyperref[sec:SISuper]{h-BN periodicity} in the SI, the h-BN periodicity and superstructure are strongly dependent on the experimental parameters, in particular the growth temperature. The exact h-BN periodicity and the Moir{\'e} pattern upon h-BN formation on Ru(0001)\cite{lu2013,goriachko2007}, due to the small lattice mismatch between $a_{\mathrm{h-BN}}$ and $a_{\mathrm{Ru(0001)}}$, is determined by the growth temperature due to a so-called a ``lock-in'' effect at that respective growth temperature\cite{martoccia2010b}. Together with the above reported additional structures, it confirms the complexity of the whole system and its dependence on minute changes of the growth parameters.

%#####################################################
\section*{Discussion and summary} 
%#####################################################
In summary, we investigated the growth of h-BN on a Ru(0001) substrate using helium atom scattering. Employing various growth conditions, characteristic periodic structures are measured during borazine exposure in addition to the h-BN diffraction peak as outlined in the diagram of \autoref{fig:schem_phasediag}. Between $750$ and \SI{880}{\K} a structure with $(3\times3)$ periodicity, that precedes the growth of h-BN, is observed with the minimum temperature necessary to form a h-BN overlayer being \SI{880}{\K}. Above this temperature, in addition to the emerging h-BN layer, we observe additional structures with a $(3\times3)$ superstructure followed by a $(3\times4)$ diffraction pattern, eventually disappearing and leaving a complete h-BN overlayer behind.

It is clearly evident from our observations that a precursor structure precedes the growth of h-BN at lower temperatures and an additional structure co-exists with h-BN at higher temperatures. Both are strongly dependent on the growth conditions, but always transform into a fully h-BN covered substrate at sufficiently high temperatures, thus confirming that the latter is the thermodynamically most stable structure. Our study of the structural evolution during the arrangement of h-BN from the precursor gas illustrates steps in the formation process itself and we hope to encourage future studies linking our structural information with  chemical characterisation.

We believe that these intermediate metastable structures may be present in many more systems where 2D materials are grown based on precursor-based CVD, at least at lower temperatures and for higher amounts of excess hydrogen compared to the ``ideal'' growth conditions (see \nameref{sec:2DOther} in the supplementary information). In the case studied here, they ultimately always transform into the complete 2D layer - and thus usually higher temperatures are reported as the ``ideal'' growth conditions for h-BN in the literature.
%In the case studied here, they ultimately always transform into the complete 2D layer - and thus usually higher temperatures are reported as the ``ideal'' growth conditions for h-BN in the literature.

These intermediate structures seem to have been largely overlooked so far. Possibly, because they are difficult to detect owing to experimental complications since the structural advent of 2D materials is often not investigated during the growth itself, or is only accessible \emph{ex situ}. More importantly, with increasing growth temperature the transformation to h-BN may occur so fast that they are easily missed.\cite{steiner2019}

The strong dependence regarding the emergence of these structures on temperature and exposure suggests that further uncovered ``routes'' and polymerisation steps are viable and the system may present an ideal playground to end up with different nano-structures. It further suggests that a careful tuning of the growth conditions via temperature and excess hydrogen from the precursor may provide new broadly applicable strategies for controlling the growth of specific nanostructures. Additional possibilities involve changing the substrate or the precursor gas, and hence tuning the thermochemistry of the surface-adsorbate complex which may further alter the subsequent reaction pathway. E.g. by changing the substrate, the metal-N bond strength may be tuned since one expects the bonding strength to increase as one moves from right to left in the transition metal series. We hope that the wide ranging implications for a controlled growth of 2D materials and nanostructures will stimulate a broad range of new research, understanding and application.

\section*{Conflicts of interest}
The authors declare no competing financial interest. 

\section*{Acknowledgements}
This research was funded in whole, or in part, by the Austrian Science Fund (FWF) [P29641-N36 \& P34704-N]. For the purpose of open access, the author has applied a CC BY public copyright licence to any Author Accepted Manuscript version arising from this submission. We would also like to thank W. Allison for helpful discussion regarding the interpretation of the data, Chris Pickard for additional structure calculations based on a neural network approach and Moritz Will for his advice in terms of the precursor (borazine) purchase and treatment. A. R. acknowledges funding by the Doctoral School and a scholarship of TU Graz. The authors acknowledge use of facilities at and support by the Cambridge Atom Scattering Centre (\url{https://atomscattering.phy.cam.ac.uk}) and the EPSRC award EP/T00634X/1  with the help of J. Ellis. M.S. is grateful for support from the Royal Society (URF/R/191029) and funding through the EPSRC (EP/S029834/1). This work used the ARCHER2 UK National Supercomputing Service (\url{http://www.archer2.ac.uk}) via membership of the UK's HEC Materials Chemistry Consortium, which is funded by the EPSRC (EP/R029431).

\section*{Supplemental information}
Supplementary material can be found online.

\section*{Data availability}
The datasets generated and analysed during the current study are available from the \href{https://doi.org/10.3217/n9db1-sf358}{TU Graz repository}, with the identifier \href{https://doi.org/10.3217/n9db1-sf358}{10.3217/n9db1-sf358}.\\

% --- References Section -------------------------------------
\setlength{\bibsep}{0.0pt}
%\makeatletter 
\putbib[literature]
\end{bibunit}
%\makeatother

\newpage

%#####################################################
% SI information
%#####################################################
\begin{bibunit}
%%%%%%%%%% Prefix Supplementary
\setcounter{equation}{0}
\setcounter{figure}{0}
\setcounter{table}{0}
\setcounter{section}{0}
\makeatletter

\renewcommand{\thefigure}{\arabic{figure}}
\renewcommand{\thetable}{\arabic{table}}
\renewcommand{\figurename}{Supplementary Figure}
\renewcommand{\tablename}{Supplementary Table}
\renewcommand{\refname}{Supplementary References}
\renewcommand{\figureautorefname}{Supplementary Figure}
\newcommand{\subfigureautorefname}{Supplementary Figure}
\renewcommand{\tableautorefname}{Supplementary Table}
\makeatletter
\renewcommand\section{\@startsection{section}{1}{\z@}%
                                  {-3.5ex \@plus -1ex \@minus -.2ex}%
                                  {2.3ex \@plus.2ex}%
                                  {\normalfont\large\bfseries}}
\makeatother

\twocolumn[\begin{@twocolumnfalse}
\LARGE{Supplementary Information for}\\ \textbf{\Large{Evolution of ordered nanoporous phases during h-BN growth: Controlling the route from gas-phase precursor to 2D material by \emph{in-situ} monitoring.}
}
%\newline \large{Ruckhofer \emph{et al.}}
~\\
~\\
\end{@twocolumnfalse}]

\setcounter{page}{1}

\section{Experimental Section}
\label{sec:SIExp}
\begin{figure}[htbp]
\centering
\includegraphics[width=0.48\textwidth]{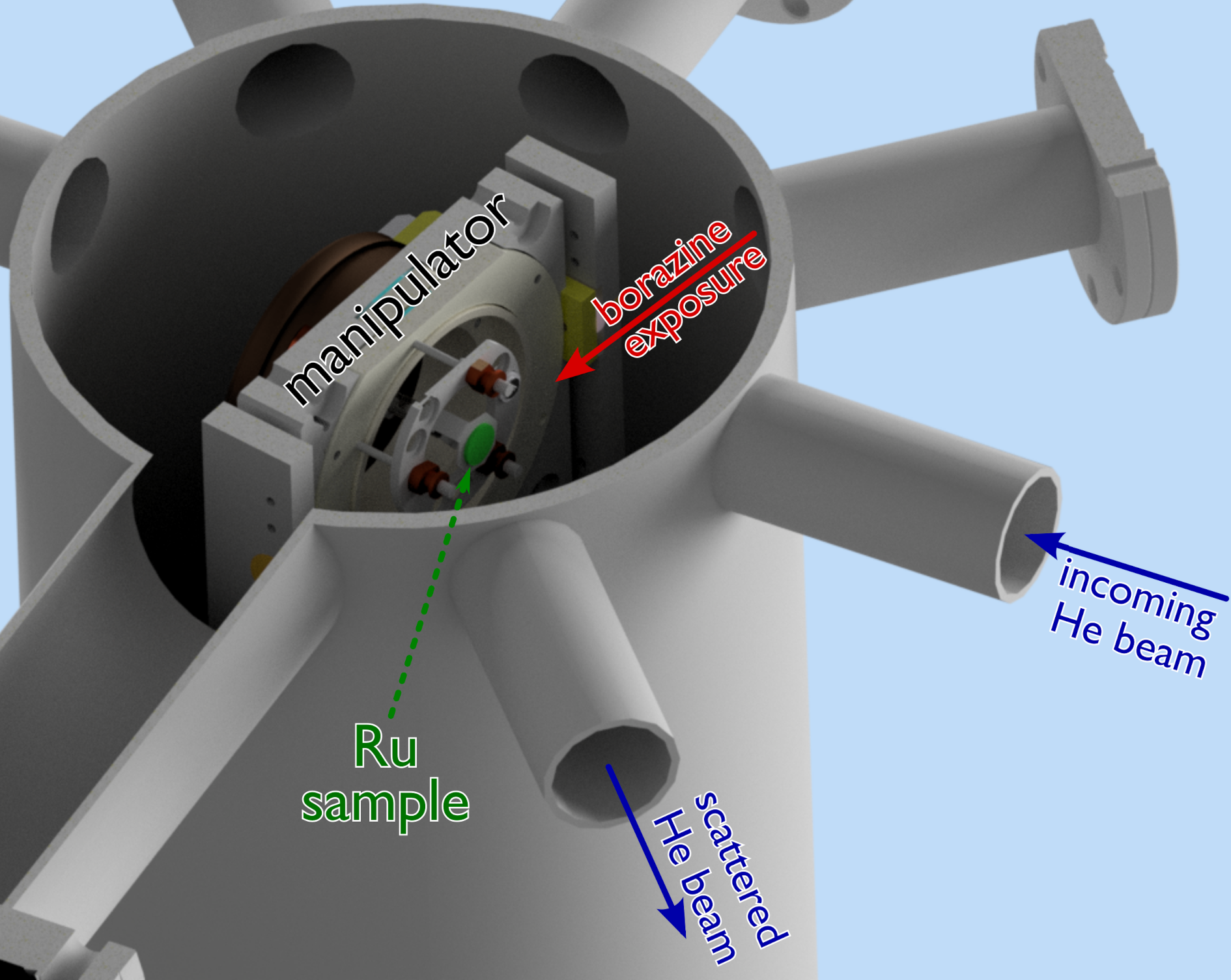}
\caption[Experimental setup]{Schematic of the experimental setup in the helium scattering chamber. The incoming He beam is scattered off the ruthenium (Ru) sample in a fixed source-detector configuration with an angle of \SI{44.4}{\degree}. The sample is mounted onto a 5-axis manipulator and can be exposed to borazine via dosing thorough a leak valve.}
\label{fig:setup}
\end{figure}

All measurements were performed with the $^3$He spin echo apparatus at the \href{https://atomscattering.phy.cam.ac.uk}{Cambridge Atom Scattering Centre}. A schematic of the scattering chamber in the experimental setup is shown in \autoref{fig:setup}. The helium beam is produced by supersonic expansion of $^3$He gas through a nozzle and enters the scattering chamber through a series of differential pumping stages. The incident helium beam is scattered off the sample, which is, together with a sample holder, mounted on a 6-axis manipulator.\cite{tamtogl2016a} Atoms travelling in a particular outgoing direction pass along the second arm of the instrument, at 44.4$^{\circ}$ total scattering angle, and are then ionised and counted in a high sensitivity mass-spectrometer detector. The incidence angle, $\vartheta_i$, with respect to the surface normal, can be varied to control the momentum transfer on scattering. 
%###########################################################################
%##########################################################################
The Ru sample can be heated radiatively and by electron-bombardment from backside of the crystal and cooled via a thermal connection to a liquid nitrogen reservoir. The entire beamline is held at high vacuum to avoid any attenuation of the helium beam, and the sample and detector chamber require ultra high vacuum levels to maintain cleanness of the sample and a low $^3$He background. The dosing was performed by backfilling the scattering chamber with borazine vapour, with borazine as provided by Katchem. To monitor the dosing rates, the chamber pressure was monitored, with typical overpressures between \SI{1e-9}{} and \SI{5e-8}{}\,mbar. At stages where borazine was not used for dosing the container was held at temperatures below \SI{0}{\degreeCelsius}.

The Ru(0001) surface was cleaned by Ar-sputtering and annealing to \SI{1300}{\K} with subsequent $\mathrm{O}_2$ treatment to not less than \SI{20}{\L} at \SI{700}{\K}. The adsorbed $\mathrm{O}_2$ was removed by repeated flashing cycles to \SI{1200}{\K}. The cleanliness of the sample was determined by helium reflectivity measurements and diffraction scans to show no features of adsorbed species. After reaching reflectivities of $\approx 23\%$ the sample was ready for the various dosing conditions. h-BN overlayers were removed by oxygen treatment at a sample temperature of \SI{900}{\K}, followed by the cleaning explained above. Borazine was supplied to the sample by backfilling the chamber through a leak-valve with typical overpressures between \SI{1e-9}{} and \SI{5e-8}{}\,mbar.

\section{Computational Methods}
\label{sec:DFTMethods}
For the DFT calculations we employed CASTEP,\cite{clark2009} a plane wave periodic boundary condition code. The plane wave basis set was truncated at an electron energy cut-off of \SI{400}{\eV} and we employ Vanderbilt ultrasoft pseudopotentials.\cite{vanderbilt1990} The Brillouin zone was sampled with a $(4\times 4 \times 1)$ Monkhorst-Pack $k$-point mesh.\cite{monkhorst1976} The Perdew Burke Ernzerhof exchange correlation functional\cite{perdew1996} was applied in combination with the Tkatchenko and Scheffler dispersion correction method.\cite{tkatchenko2009} The Ru(0001) surface was modelled by a 5-layer slab in a $(3\times3)$ supercell, and an additional \SI{15}{\angstrom} vacuum layer for separating the periodically repeated supercells in the $z$-direction. Positions of the atoms in the adsorbate and in the top three layers of the Ru substrate were left fully unconstrained. For the structural optimisations, the force tolerance was set to $0.05\,\mbox{eV} / \mbox{\AA}$. 

The adsorption energies $E_{\mathrm{ads}}$ are defined to be: 
$$E_{\mathrm{ads}}= E_{\mathrm{tot}}(x+n\,y) - E_{\mathrm{tot}}(x) - nE_{\mathrm{tot}}(y)$$  
where $E_{\mathrm{tot}}(x+n\,y)$ is the total energy of the system, $E_{\mathrm{tot}}(x)$ is the energy of the substrate, $E_{\mathrm{tot}}(y)$ is the energy of the adsorbate and $n$ is the number of adsorbed molecules. The more negative $E_{\mathrm{ads}}$, the more thermodynamically favourable it is for the species to adsorb.

In order to compare the intermediate structures with a different number of atoms we calculate the binding energy $E_{\mathrm{bin}}$ relative to Ru(0001) + 3 borazine molecules (3 borazine molecules are needed to form h-BN on a $(3\times3)$ cell) by  appropriately adding or subtracting the energy of H$_2$ and borazine in the gas phase, to preserve stoichiometry:
{\small
$$ E_{\mathrm{bin}} = E_{\mathrm{tot}} + \dfrac{n_\mathrm{H}}{2} E_{\mathrm{tot}}(\mathrm{H}_2) + n_{\mathrm{BZ}} E_{\mathrm{tot}}(\mathrm{BZ}) - E_{\mathrm{tot}}(\mathrm{Ru}) - E_{\mathrm{tot}}(3\mathrm{BZ}) $$}
where $E_{\mathrm{tot}}$ is the total energy of the system, $E_{\mathrm{tot}}(\mathrm{H}_2)$ and $E_{\mathrm{tot}}(\mathrm{BZ})$ are the energies of H$_2$ and borazine which remain in the gas phase, respectively and $E_{\mathrm{tot}}(\mathrm{Ru})$ and $E_{\mathrm{tot}}(3\mathrm{BZ})$ are the total energies of pristine Ru(0001) and 3 borazine molecules in the gas phase. The more negative $E_{\mathrm{bin}}$, the stronger the binding and it becomes thermodynamically more favourable for the species to form.

\section{Supplementary DFT calculations}
\label{sec:SIDFT}
\begin{figure}[htbp]
\centering
\includegraphics[width=0.48\textwidth]{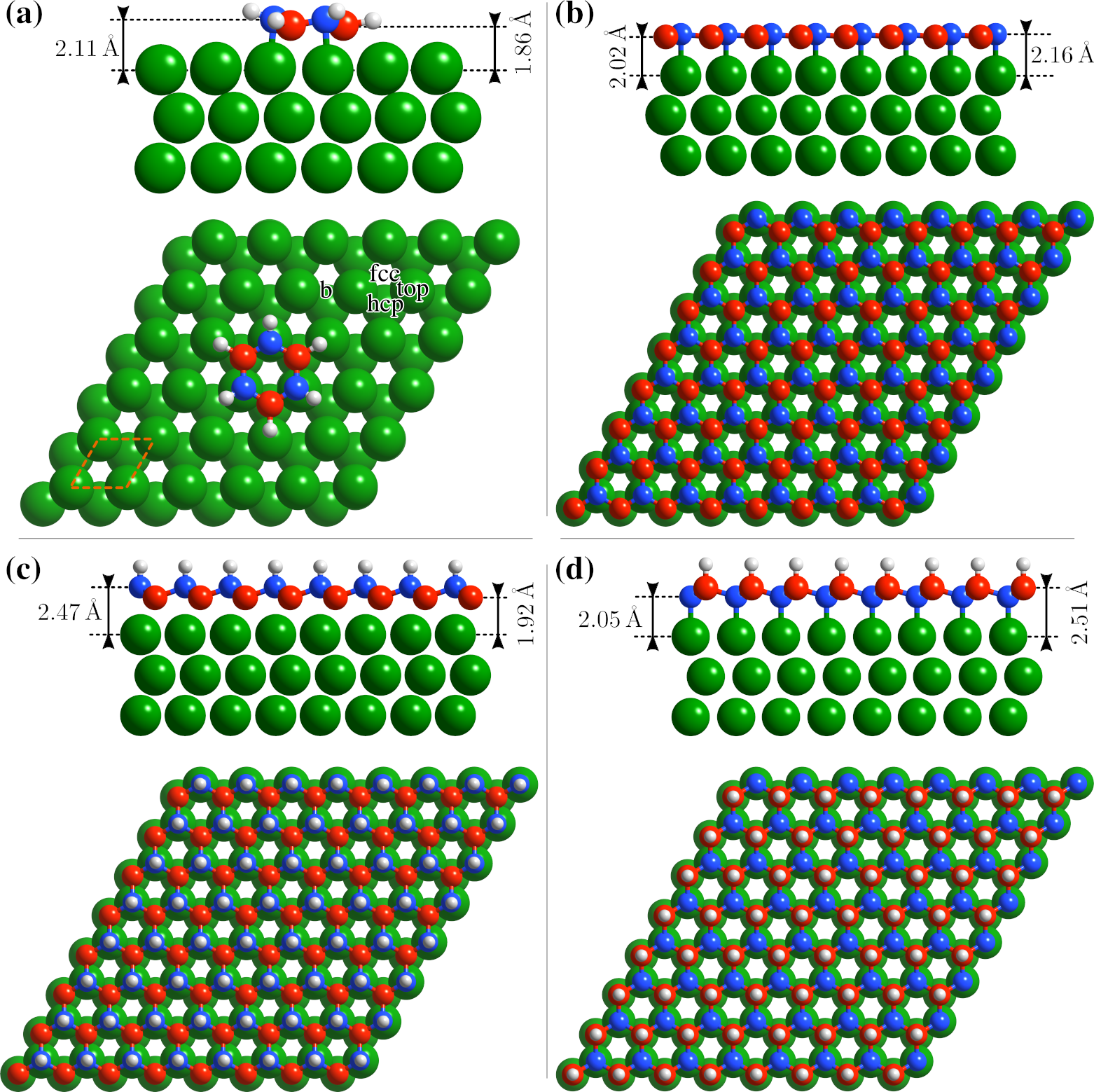}
\caption[Diffraction pattern]{(a) depicts the outcome of vdW-corrected DFT calculations for one borazine molecule (B$_3$N$_3$H$_6$) on Ru(0001), with the most favourable adsorption site being the fcc position. All possible adsorption sites are given with their according labels. (b) shows the top and side-view of h-BN on Ru(0001) with the corresponding interatomic distances. In (c) and (d) the optimised geometry for three partially dehydrogenated borazine molecules is illustrated, which essentially forms a hydrogenated version of h-BN/Ru(0001). Here (c) represents a ``local'' minimum as also seen from \autoref{tab:DFT_hBN}. The hydrogen atoms stick out of the surface, yielding a high corrugation and hence the buckling will be different once the structure is dehydrogenated, such as for a complete h-BN layer with its corrugation reflecting the Moiré pattern.}
\label{fig:DFT1}
\end{figure}

The energetically most favourable adsorption site for a single intact borazine molecule per $(3\times3)$ supercell according to DFT calculations is shown in \autoref{fig:DFT1}(a). The adsorption sites (top, hcp, fcc, b) are given relative to the centre of the borazine molecule. 
%#################################################################
%#################################################################
In addition to the calculations for one borazine molecule given in the main text we show the results for the intact and partly dehydrogenated molecule with an initial rotation of 0$^\circ$ in \autoref{tab:1BZ_ads_0deg}. When comparing the results we now see that in this case the hcp site is energetically most favourable with an adsorption energy of E$_\mathrm{ads}$=\SI{-8.85}{\eV}. If the borazine molecule is initially placed on a bridge site it undergoes a transition to the hcp position. Still, the results for the 60$^\circ$ rotation are energetically more favourable by $\approx\SI{0.1}{\eV}$.\\
\autoref{fig:DFT1}(c,d) shows that the outcome for calculations considering three partially dehydrogenated borazine molecules on Ru(0001), results in a structure similar to h-BN, except for the fact that the H atoms remain attached to the boron/nitrogen atoms. For comparison, \autoref{fig:DFT1}(b) depicts the optimised structure for h-BN/Ru(0001).\\
\autoref{tab:DFT_hBN} illustrates that hydrogenation of the h-BN overlayer becomes thermodynamically unfavourable due to the correction with respect to molecular hydrogen in the gas phase and the high binding energy of the latter. The result is in line with hydrogenation experiments of metal supported h-BN, where  atomic hydrogen exposure is required in order to facilitate the hydrogenation.\cite{spaethSI} Interestingly, in contrast to h-BN/Ni(111),\cite{spaethSI} H adsorption on top of the N-site is slightly more favourable than on top of the boron site for h-BN/Ru(0001) as can be seen from the adsorption energy per hydrogen atom.
%#################################################
\begin{table}[htbp]
\centering
\caption{DFT calculations for the adsorption structures of the borazine precursor on Ru(0001), based on a $(3\times3)$ supercell with one molecule per cell. The results are shown for an intact (B$_3$N$_3$H$_6$) and partially dehydrogenated (B$_3$N$_3$H$_3$) adsorbate, considering various initial adsorption sites and a rotation of \SI{0}{\degree}. The adsorption energies E$_\mathrm{ads}$ are given for the final optimised adsorption site and $\Delta$E is the difference with respect to the minimum energy configuration of the system with the same dehydrogenation state.}
\resizebox{.47\textwidth}{!}{
\begin{tabular}{ c | c | c || c | c | c }
	\toprule
	\multicolumn{3}{c||}{\textbf{B$_3$N$_3$H$_6$} } & \multicolumn{3}{c}{\textbf{B$_3$N$_3$H$_3$}} \\[1mm]
	\midrule
	\rule{0mm}{5mm}
	Site & E$_\mathrm{ads}$ (eV)  & $\Delta$E (eV)   & Site & E$_\mathrm{ads}$ (eV)  & $\Delta$E (eV)\\ [1mm]
	\midrule
	\rule{0mm}{5mm}
	 fcc                & -1.28 & 2.80 &  fcc               & -5.96  & 2.99 \\
	 top                & -1.21 & 2.87 &  top               & -6.60  & 2.35  \\
     b$\rightarrow$hcp  & -3.99 & 0.08 &  b$\rightarrow$hcp & -8.85  & 0.11 \\
	 hcp                & -3.99 & 0.08 &  hcp               & -8.85  & 0.11 \\[1mm]
	\bottomrule
	\end{tabular}}
\label{tab:1BZ_ads_0deg}	
\end{table}
%################################################# 

From the side view in \autoref{fig:DFT1}(c,d) it becomes evident that the closest atom to the Ru substrate and the bond length change, depending whether nitrogen or boron remain hydrogenated. In \autoref{fig:DFT1}(d) the hydrogen atoms appear to “pull” the boron away from the surface by \SI{0.5}{\angstrom} and the sp$^2$ hybridised bonds to nitrogen gain more sp$^3$ character. Therefore the boron atom moves away from the surface to optimise these bonds, forming a tetrahedral (bond angle \SI{106}{\degree}). Likewise the nitrogen binds to the Ru orbitals, thus moving closer to the surface. If hydrogen desorbs from this structure pure h-BN is formed, as seen in \autoref{fig:DFT1}(b). The boron-nitrogen bonds become stronger and therefore boron moves \SI{0.5}{\angstrom} towards the Ru, to be in the same plane as the nitrogen. In addition the nitrogen orbitals are populated from the boron and the nitrogen-Ru interaction is weakened, resulting in a movement of the nitrogen atoms \SI{0.11}{\angstrom} away from the Ru surface. For pure h-BN on Ru, the boron atoms are positioned only slightly lower than the nitrogen atoms (\SI{0.14}{\angstrom}). This may reflect the gain in stability from Ru-B bonding when boron is moved slightly into the hole site, compared to maintaining perfect sp$^2$ hybridised bonds.

%##########################################################################
\begin{figure*}[htp]
\centering
\includegraphics[width=0.8\linewidth]{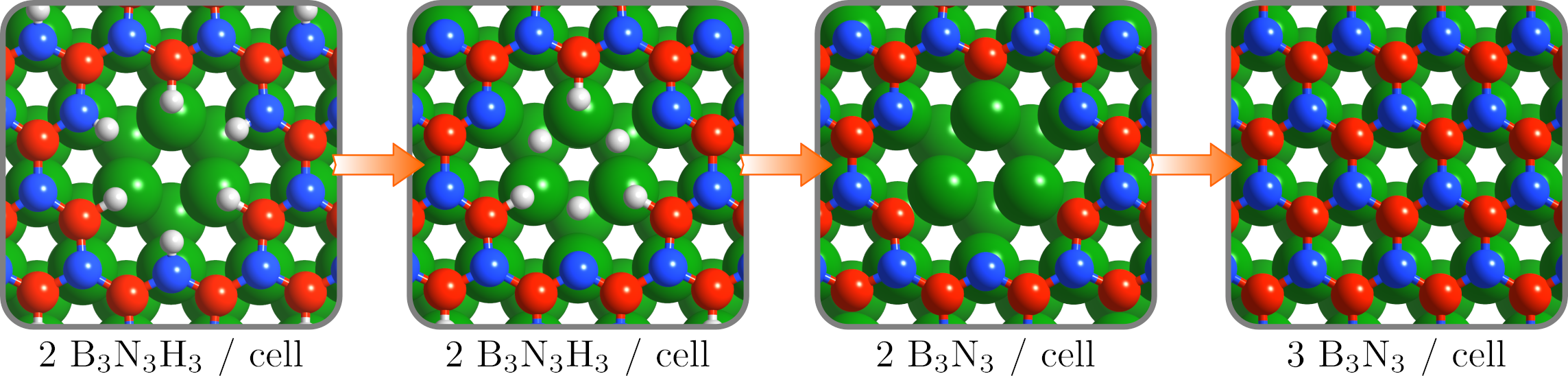}
\caption{Schematic of the possible route from the partially dehydrogenated precursor structure to h-BN via several intermediate structures based on vdW-corrected DFT calculations. As noted in the text, the precursor structure is already quite close to the binding energy for the complete h-BN layer. In contrast to the DFT calculations, where entropy contributions were not considered, additional dehdrogenation and bond breaking may occur due to the high experimental temperatures.}
\label{fig:hBNRoute}
\end{figure*}
%##########################################################################
{\small
%#################################################
\begin{table}[htbp]
\centering
\caption{DFT calculations for different configurations/structures with their respective energies.}
\begin{tabularx}{0.94\linewidth}{C{2.0cm} | C{2.9cm} | C{1.8cm}}
	\toprule
	Configuration  &  Adsorption energy per H atom (eV)  &  Adsorption site \\
	\midrule
	hydrogenated h-BN/Ru &  0.84 & H on top of B \\ [1mm]  %7.59/8
	hydrogenated h-BN/Ru &  0.81 & H on top of N \\ [1mm]  %7.31/8
	\hline
	\bottomrule
\end{tabularx}
\\[0.2cm]
\begin{tabularx}{0.94\linewidth}{C{2.0cm} | C{2.9cm} | C{1.8cm}}
	\toprule
	%\rule{0mm}{5mm}
	%\multirow{2}{*}{Configuration}  &  Binding  & \multirow{2}{*}{Stacking} \\
	%  & energy (eV) &  \\
	Configuration  &  Physisorption energy (eV)  & Stacking \\
	\midrule
	%h-BN on Ru  & -6.74 & - \\ [1mm]
	h-BN on h-BN/Ru  & -2.09 & AB \\ [1mm]
	Borazine on h-BN/Ru & -0.73 & AB\\ [1mm]
	\bottomrule 
\end{tabularx}
\label{tab:DFT_hBN}
\end{table}
%################################################# 
}

As mentioned in the main text, we considered also borazine adsorption on top of h-BN/Ru as well as the formation of bi-layer h-BN. The physisorption energies are shown in the lower part of \autoref{tab:DFT_hBN}, illustrating that both are thermodynamically favourable with a stronger physisorption energy for a second h-BN layer on top of h-BN/Ru. On the other hand, the corresponding binding energy for a single complete h-BN layer is \SI{-6.74}{\eV} upon formation from 3 borazine molecules per supercell on Ru(0001). In the following we consider a possible route to the complete h-BN layer starting from the precursor structure as described in the main paper,

\autoref{fig:hBNRoute} shows the route through various steps based on DFT calculations. The first (precursor) structure is strongly bound with a binding energy $E_{\mathrm{bind}}$ (see \hyperref[sec:DFTMethods]{Computational methods}) of \SI{-6.28}{\eV} in relation to the bare Ru surface and the molecules in the gas phase. The next step towards h-BN formation, involves dehydrogenation. The calculations show that if the three hydrogen atoms are detached only from the boron atoms they eventually reattach to the same boron sites. Therefore, initially the hydrogen atoms are detached from the nitrogen atoms which adsorb on the Ru substrate within the nanopores, yielding a binding energy of \SI{-3.27}{\eV}. At sufficient surface temperature eventually all hydrogen atoms will desorb from the surface yielding the third structure with a less favourable energy of \SI{1.26}{\eV}. If now the nanopore is filled with one additional borazine atom, h-BN is formed yielding the lowest binding energy (\SI{-6.74}{\eV}). Therefore we conclude that the first structure is nearly as stable as h-BN and that on the route to h-BN several energy barriers have to be overcome. It should be mentioned however, that the calculations were performed at \SI{0}{\K} and that no entropy contributions were considered.

\section{Supplementary diffraction scans}
\label{sec:SIDiff1}
\begin{figure}[htbp]
\centering
\includegraphics[width=0.48\textwidth]{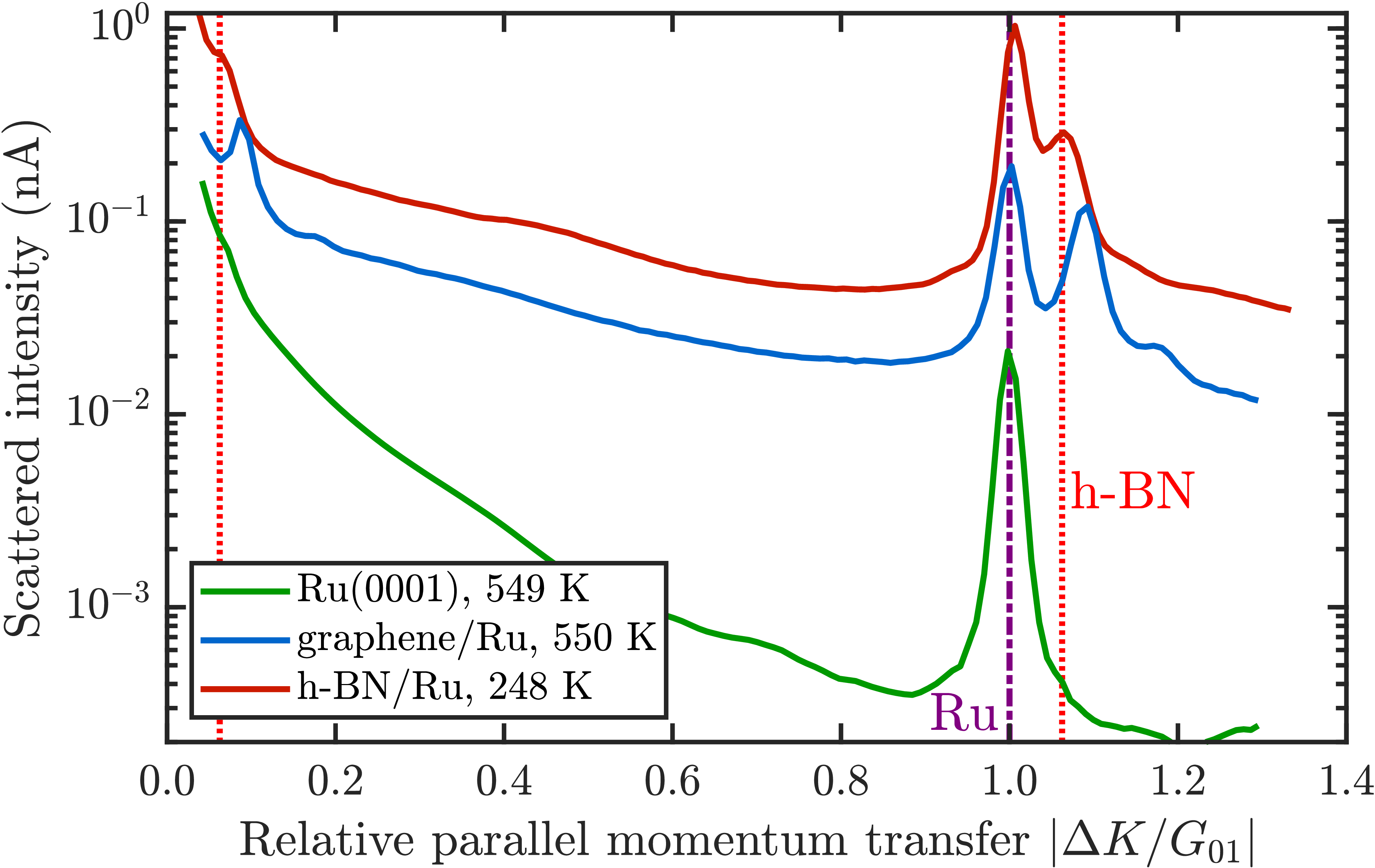}
\caption[Diffraction pattern]{Angular diffraction patterns using helium scattering. Comparison of the diffraction scans of the clean Ru(0001) surface (green), graphene on Ru (blue) and h-BN on Ru. The purple dash-dotted line indicates the first order diffraction position and the red dotted lines the h-BN reconstruction peaks.}
\label{fig:graphene_hBN}
\end{figure}

h-BN, sometimes also called ``white graphene'', typically forms a Moir\'{e} pattern on the surfaces of reactive transition metals such as Rh(111) or Ru(0001), as mentioned in the main text. The two-dimensional h-BN layer on such surfaces exhibits periodic nanometric structures, often called ``nanomesh'', with areas which are elevated from the surface, and areas closer to the surface. In \autoref{fig:graphene_hBN} the characteristic diffraction pattern of the clean Ru sample (green) is compared to two overlayers on the same substrate. 
%###################################################################
%###################################################################

The scans of the single layer graphene and h-BN covered Ru show additional peaks close to the specular and first order Ru diffraction peaks. The blue curve depicts the scattering result for a graphene monolayer on Ru which has been studied extensively in earlier works.\cite{borca2010SI,maccariello2015} The graphene layer was grown by heating the Ru crystal to \SI{1250}{\K} for several minutes. Leaving the crystal at such high temperatures brings the carbon out of the bulk which then forms the honeycomb single layer graphene sheet. Graphene forms a (12-on-11) superstructure in which a $(12\times12)$ supercell of graphene coincides with a $(11\times11)$ supercell of ruthenium, giving rise to additional diffraction peaks at $|\Delta K/G_{01}| =  1/11 \equiv 0.09$ and $|\Delta K/G_{01}| =  12/11 \equiv 1.09$.

The diffraction pattern for h-BN on a Ru substrate is depicted in red in \autoref{fig:graphene_hBN}. Earlier works indicate that h-BN forms a (13-on-12) superstructure which can be identified by position of the diffraction peaks.\cite{martoccia2010SI} Indeed the feature originating from the h-BN nanomesh to the right of the Ru diffraction peak shifts to smaller values of $|\Delta K/G_{01}|$ with respect to graphene, giving rise to a bigger supercell. 

In \autoref{fig:graphene_hBN} the scans for pure Ru and graphene were performed at a sample temperature of $T=\SI{550}{\K}$ while the scan of h-BN was taken at \SI{248}{\K}. Due to thermal expansion it gives rise to a deviation of the position of the first order substrate (Ru) peak for the h-BN scan compared to the other two measurements as indicated by the purple line. In all scans the specular peak (at $|\Delta K/G_{01}| =  0$) was cut off due the high intensity and the first order diffraction peak of the Ru surface corresponds to $|\Delta K/G_{01}| = 1$.
%###################################################################
\begin{figure}[htbp]
\centering
\includegraphics[width=0.48\textwidth]{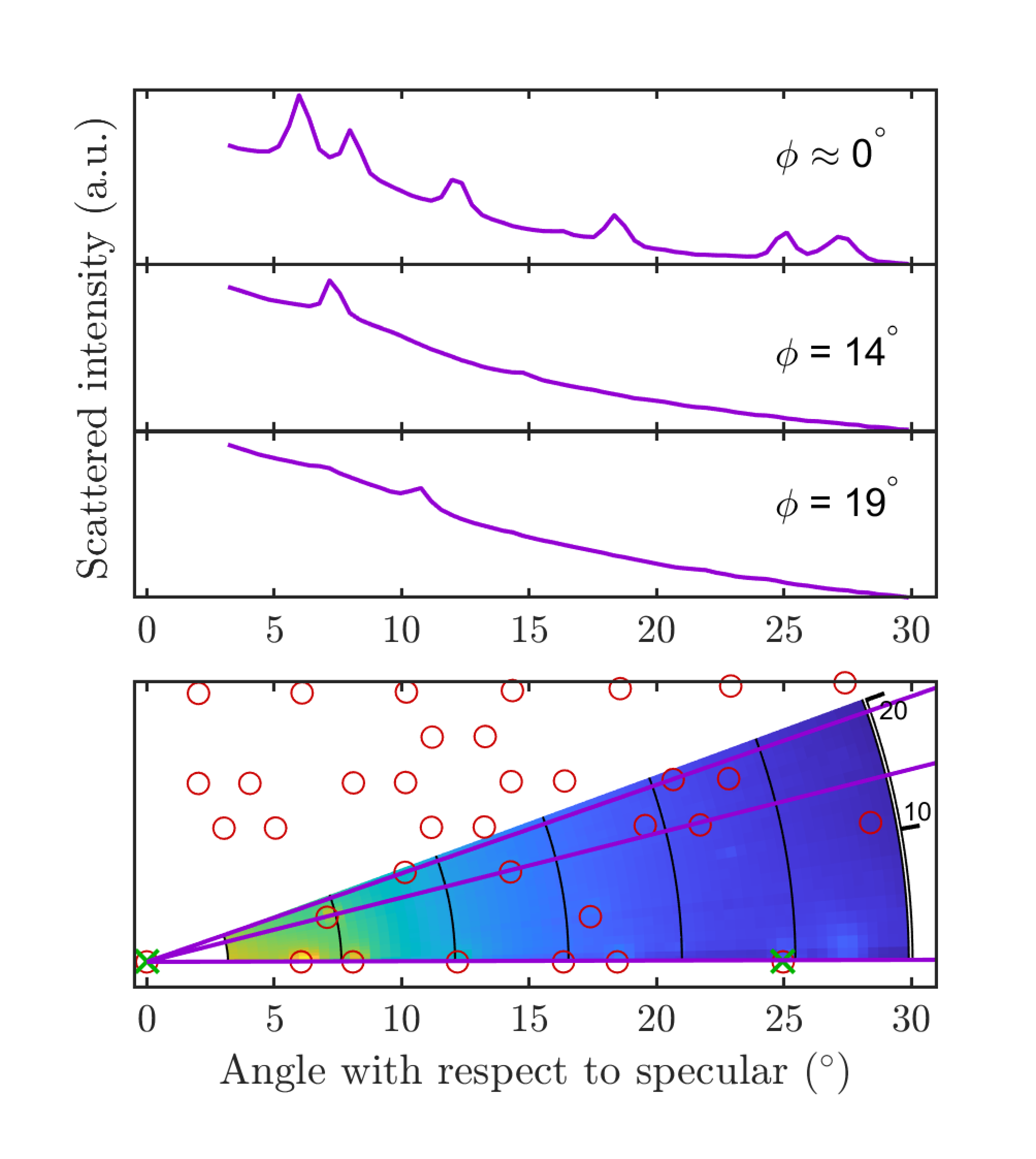}
\caption[2-dimensional scan of the $(3\times4)$ structure]{2-dimensional scan of the $(3\times4)$ structure of the adsorbed borazine molecules on the Ru surface. The polar plot consists of 22 individual logarithmic diffraction scans at various azimuthal orientations $\varphi$ and a surface temperature of \SI{300}{\K}. The red circles indicate the calculated scattering positions for a $(3\times4)$ superstructure while the green crosses mark the Ru diffraction positions. Three exemplary scans at the top are drawn to elucidate the diffraction peak positions in dependence of $\varphi$. }
\label{fig:2dscan}
\end{figure}
%###################################################################

In addition \autoref{fig:graphene_hBN} clearly shows that the background intensity between the Ru diffraction peaks is much lower, indicating less inelastic scattering and fewer diffuse scattering when probing the clean Ru crystal. In both diffraction scans of h-BN and graphene the background increases by two orders of magnitude due to the increase of diffuse scattering. In addition, adlayers change the corrugation at the surface which is observed by the He atoms. X-ray studies showed that the peak-to-peak corrugation height of graphene is $(0.82\pm 0.15)$, whereas for the uppermost Ru-atomic layer it is $(0.19\pm 0.02)$.\cite{martoccia2010c}

Performing a two-dimensional (2D) scan confirms that the diffraction peaks in the 1D angular diffraction scan of Figure \ref{fig:superstruct}(b) in the main text are correctly assigned to a $(3\times4)$ periodicity and cannot be explained as a subset of another periodicity or as domains with different rotations. Therefore we performed diffraction scans at various azimuthal orientations, since the BN$_{\RN{2}}$ structure has very distinct diffraction peaks in the high symmetry direction as well as along other azimuthal orientations. By rotation of the azimuthal angle of the sample a 2D-plot in reciprocal space can be created (see \autoref{fig:2dscan}). The green cross marks the Ru diffraction peak while the red circles indicate the calculated positions of the $(3\times4)$ structure peaks. In the top panel three exemplary diffraction scans at specific azimuthal angles $\varphi$ are depicted. Small angles close to the specular peak are not shown due to their high intensity in all scans. The identification of the peaks verifies the assumption that the $(3\times4)$ structure is present in addition to the h-BN layer on the surface and cannot be explained e.g. as being part of another superstructure or rotated domains of a $(3\times3)$ structure.
%#####################################################
\begin{figure}[htbp]
\centering
\includegraphics[width=0.46\textwidth]{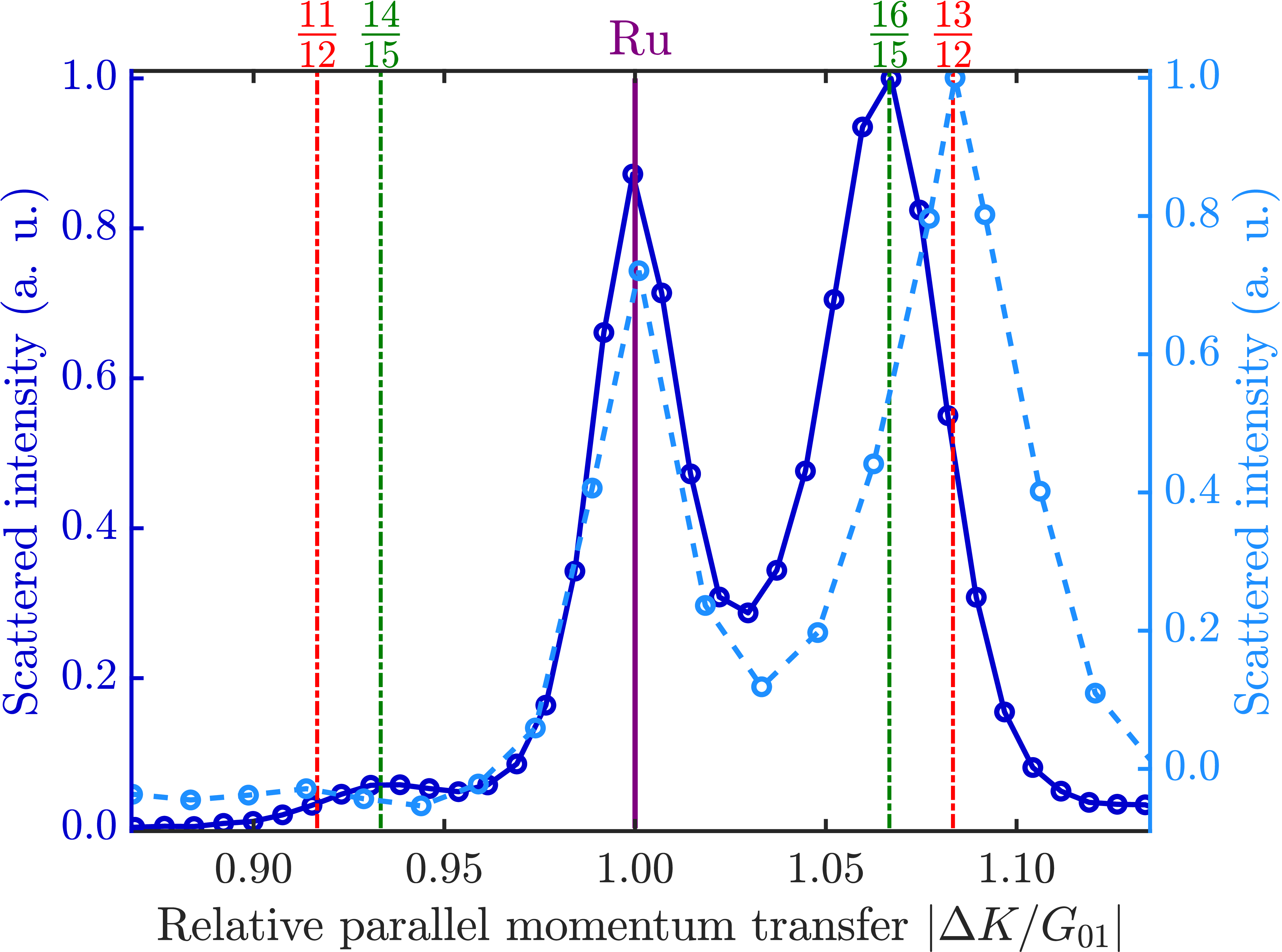}
\caption[h-BN peak position]{Diffraction scans of the h-BN periodicity illustrate that the exact superstructure of the overlayer depends on the growth temperature, with the blue scan for h-BN grown at \SI{1020}{\K} and the dashed cyan curve for h-BN grown at \SI{900}{\K}. The h-BN peaks at the right-hand side of the Ru peak at $|\Delta K/G_{01}| =  1$ show, that h-BN adopts a larger superstructure with increasing growth temperature. Due to the decreasing lattice mismatch the overlayer adopts a $16/15$ ratio versus a $13/12$ ratio at \SI{900}{\K}. The ``lock-in'' effect (see text) is confirmed by the small (substrate) reconstruction peaks on the left-hand side. For better identification of the peaks a linear background was subtracted from the untreated data and the sample was subsequently cooled down to room temperature for the duration of the scan.}
\label{fig:fractions}
\end{figure}
%#####################################################

\section{h-BN periodicity and reconstruction}\label{sec:SISuper}
The h-BN periodicity and superstructure are strongly dependent on the experimental parameters, in particular the growth temperature. It is well known that h-BN forms a Moir{\'e} pattern on the Ru(0001) surface due to the small lattice mismatch between $a_{\mathrm{h-BN}}=\SI{2.505}{\angstrom}$ and $a_{\mathrm{Ru(0001)}}=\SI{2.706}{\angstrom}$.\cite{paszkowicz2002,arblaster2013} At room temperature, such a mismatch results in a superstructure where 13 unit cells of h-BN coincide with 12 unit cells of Ru: $(13\times13)$ on $(12\times12)$. On the other hand, previous studies on a similar substrate showed that the h-BN overlayer and the substrate lock in at the temperature during the growth with the strong interlayer bonding causing the superstructure ratio to remain constant after cooling back down.\cite{martoccia2010b}

We show that the same holds for different growth temperatures of h-BN on Ru(0001). Detailed diffraction scans around the h-BN $(01)$-peak in \autoref{fig:fractions} illustrate that for a h-BN synthesis at \SI{1020}{\K} (blue curve), the h-BN peak at $|\Delta K/G_{01}| =  1.067$ fits a superstructure ratio of $16/15$ perfectly, as shown by the green vertical dash-dotted line. Upon growing the h-BN overlayer at a lower temperature of \SI{900}{\K} (cyan curve) the h-BN peak appears at a ratio of $13/12$. The small peaks to the left of the first order Ru peak in \autoref{fig:fractions} originate from the surface reconstruction with a $14/15$ and $11/12$ ratio, respectively. These reconstruction peaks can only arise if the system exhibits a true commensurate superstructure.\cite{martoccia2010b,leconte2020}

Our HAS measurements show a strong temperature dependence and thus a strong ``lock-in'' effect, as further discussed below. Compared to X-ray diffraction where a commensurate 14-on-13 superstructure was reported,\cite{martoccia2010} we see that only h-BN growth at lower temperature (\SI{900}{\K} with a borazine exposure of \SI{15}{\L}) followed by a slow subsequent cooling provides a 13 over 12 superstructure, similar to previous studies.\cite{goriachko2007} After all, compared to the h-BN/Rh(111) system,\cite{laskowski2008} the bonding strength of the N-atoms to the Ru substrate is predicted to increase and thus one expects a stronger ``lock-in'' effect on Ru as observed above. Moreover, due to HAS being strictly surface sensitive, our results can be interpreted as scattering that stems solely from the h-BN nanomesh while other methods may contain contributions from the substrate structure. E.g, a coincidental overlay of the flat h-BN monolayer on a completely flat Ru substrate would not give rise to a diffraction pattern as shown in \autoref{fig:superstruct}(a) and \autoref{fig:fractions}. Together with the above reported additional structures, it confirms the complexity of the whole system and its dependence on minute changes of the growth parameters.

Looking at the thermal expansion coefficients of bulk h-BN and the Ru(0001) surface gives a rough estimation for the temperature at which the 13/12 superstructure is favourable. The thermal expansion of bulk h-BN\cite{pease1952} and the Ru surface\cite{ferrari2010} are given by:
\begin{subequations}
\begin{align}
a_{\mathrm{h-BN}} &= 2.505-7.42\times 10^{-6}\cdot (T-297) \nonumber \\
                  & \quad +4.79\times 10^{-9}\cdot (T-297)^2  \\
a_{\mathrm{Ru}}   &= 2.706+9.22\times 10^{-6}\cdot (T-293)
\end{align}
\end{subequations}
Here the lattice constant for Ru $a_{\mathrm{Ru}}=\SI{2.706}{\angstrom}$ was taken for a surface temperature of \SI{293}{\K}, with \SI{297}{\K} for $a_{\mathrm{h-BN}}$, hence the subtraction of these values.
%###################################################################
\begin{figure}[htbp]
\centering
\includegraphics[width=0.48\textwidth]{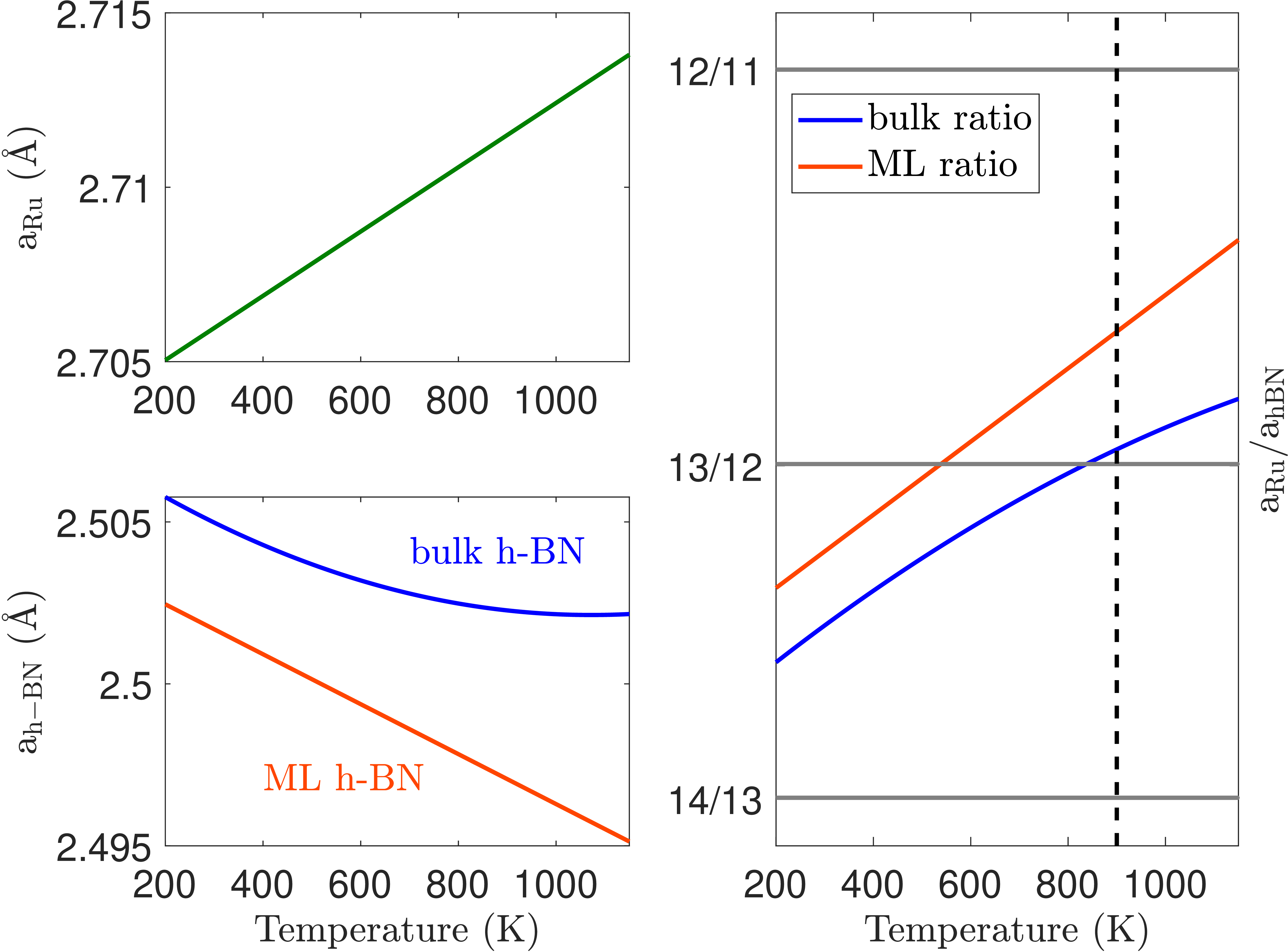}
\caption[Thermal expansion]{Thermal expansion for the Ru surface (upper left panel) and bulk h-BN as well as monolayer (ML) h-BN (lower left panel). The ratio of the Ru surface lattice constant and the h-BN lattice constant versus temperature (right panel) provides an estimate of the expected superstructure. The grey horizontal lines depict the respective superstructure ratios while the dashed vertical line indicates a growth temperature of \SI{900}{\K}.}
\label{fig:therm_exp}
\end{figure}
%###################################################################

The Ru thermal expansion is depicted in the upper left panel of \autoref{fig:therm_exp}, while the slope of bulk h-BN is shown as a blue line in the lower left panel. In addition, the thermal expansion for a single monolayer (ML) of h-BN as calculated by Thomas \emph{et al.}\cite{thomas2015} is drawn in orange. Taking the ratio of the values for h-BN and Ru then yields the expected superstructure at a given surface temperature, as shown in the right panel of \autoref{fig:therm_exp}. The expected fraction of 13/12 nicely fits the value of \SI{900}{\K} when using the bulk value of the thermal expansion.

\section{Supplementary discussion}
\label{sec:SIDiscus}
%In the following we discuss further scenarios of the $\mbox{BN}_{\RN{2}}$ structure and provide a short outlook for intermediate structures on other 2D materials. As mentioned in the main text, the surface temperature strongly influences the kinetics and thus the duration and appearance of the additional superstructures. Interestingly, after borazine exposure is stopped and stable h-BN is formed on the surface, the additional structures may even reappear upon heating to temperatures above \SI{1100}{\K} further confirming the strong dependence on the growth route and parameters.
In the following we discuss further scenarios of the $\mbox{BN}_{\RN{2}}$ structure. As mentioned in the main text, the surface temperature strongly influences the kinetics and thus the duration and appearance of the additional superstructures. %Interestingly, after borazine exposure is stopped and stable h-BN is formed on the surface, the additional structures may even reappear upon heating to temperatures above \SI{1100}{\K} further confirming the strong dependence on the growth route and parameters.
%\subsection{Discussion of the \texorpdfstring{$\mbox{BN}_{\RN{2}}$} structure}
At temperatures above \SI{1000}{\K} the $(3\times4)$ structure ($\mbox{BN}_{\RN{2}}$) slowly vanishes (see Figure \ref{fig:phasediag} in the main text) which leads to the assumption that either strongly bound atoms/molecules desorb into the gas phase or convert into another structure. As mentioned earlier the dehydrogenation of borazine already starts at lower temperatures\cite{orlando2012SI} leading to the assumption that the adsorbed species on Ru(0001) are at least partly dehydrogenated.

In the following we provide several scenarios for the origin of the $(3\times4)$ structure and discuss their plausibilities.
The results could be interpreted as if borazine converts upon adsorption to both h-BN and a $(3\times3)$ structure (BN$_{\RN{1}}$). However, given the results which are reported in the main paper, it is clear that borazine only adsorbs in a $(3\times3)$ superstructure, and at \SI{880}{\K} a (relatively fast) conversion to h-BN occurs. The h-BN and BN$_{\RN{1}}$ structure grow together until the BN$_{\RN{1}}$ reservoir is depleted, and no more h-BN is created. At this point we can conclude that the $(3\times4)$ (BN$_{\RN{2}}$) is not a precursor to h-BN and is also not converted from the BN$_{\RN{1}}$ structure. Since the $(3\times3)$ peaks degrade completely, the rise of the BN$_{\RN{2}}$ structure does not compete with the conversion of the BN$_{\RN{1}}$ structure to h-BN.

When looking at Figure \ref{fig:phasediag} in the main paper one might also think that after the BN$_{\RN{1}}$ structure vanishes and the h-BN peak saturates, that the h-BN monolayer is complete and the additional borazine exposure gives rise to a second layer being formed. This layer could consist of partly dehydrogenated borazine forming a periodic structure on top of the existing h-BN layer. According to literature, the CVD process for h-BN growth is usually considered to be self-terminating after a single layer, while some works also showed that multilayers are formed,\cite{kidambi2014SI} however, typically these require different growth approaches.\cite{tonkikh2016,jang2016,guo2012SI,sutter2013} As described in the main paper, from our experimental observations we can rule out the behaviour of multilayer h-BN growth and ascribed the BN$_{\RN{2}}$ structure to a second chemisorbed layer on top of h-BN.

Another possible scenario would be the growth of a superstructure in-between the already grown h-BN islands. As mentioned in the main manuscript an earlier work investigated the CVD growth of h-BN on Ir(111) and identified a $(6\times2)$ superstructure in-between the h-BN islands.\cite{petrovic2017SI} A similar behaviour could lead to the formation of a $(3\times4)$ structure in-between the h-BN islands on Ru. This intermediate structure eventually upon further borazine exposure converts into h-BN which connects the previously formed h-BN islands. However the areas which formed under this condition are less stable since they convert back to a $(3\times4)$ structure upon heating of the sample (see phenomenological cycle equation in the main paper). Upon further annealing of the surface the structures in-between the stable h-BN islands eventually desorb from the surface leaving behind some h-BN islands.
 %##################################################################
 \begin{figure}[htbp]
 \centering
 \includegraphics[width=0.42\textwidth]{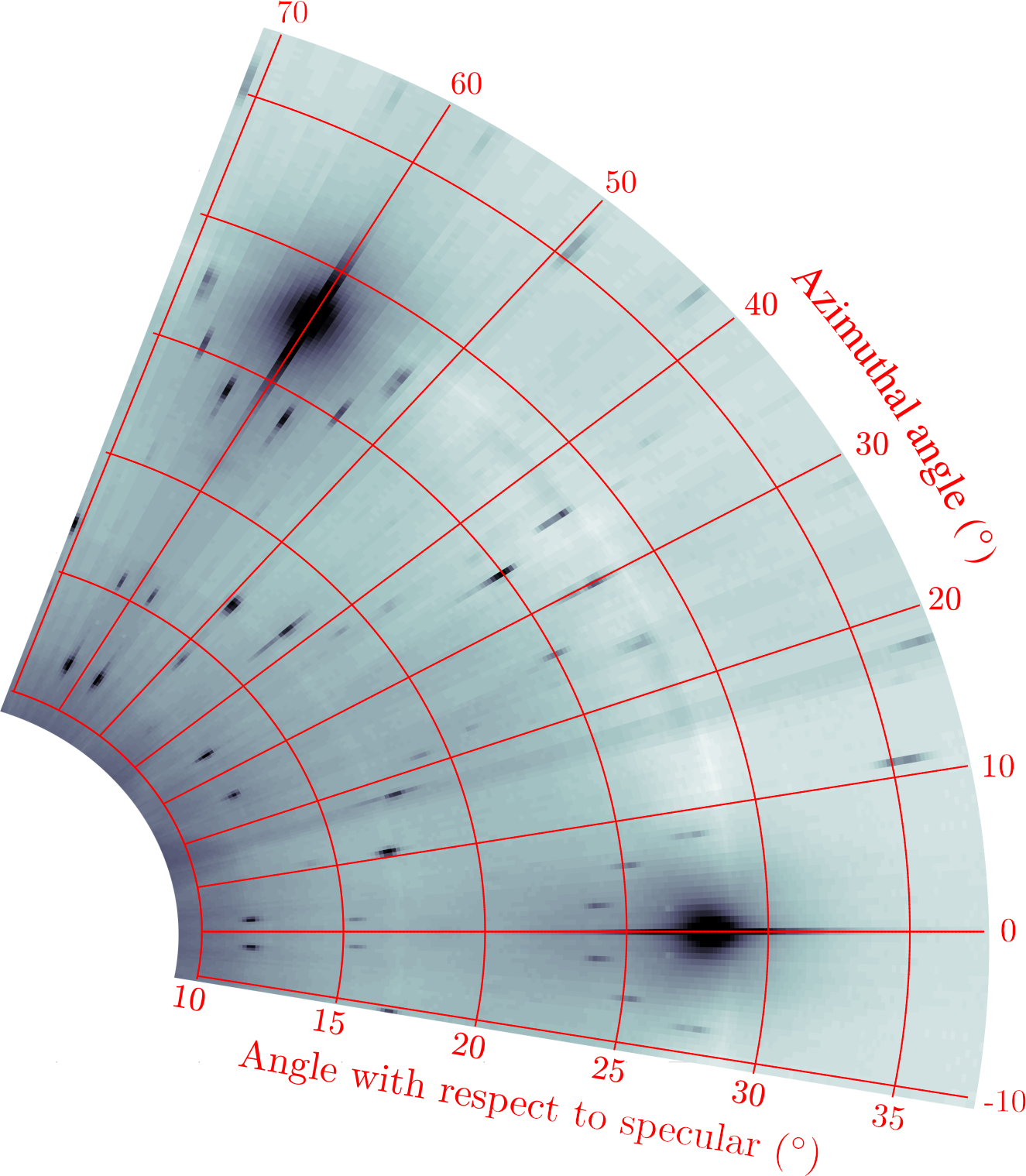}
 \caption{A 2-dimensional diffraction scan of CVD-grown graphene on Ni(111) at temperatures below the ``ideal'' growth temperatures, reveals additional diffraction peaks which transform only upon heating to \SI{730}{\K} into the $(1\times1)$ graphene/Ni(111) structure.}
 \label{fig:2DGr}
 \end{figure}
 %##################################################################

 \section{Outlook for other 2D materials}
 \label{sec:2DOther}
 We hope that our initial findings encourage future investigations to give insight on the peculiar intermediate structures during the CVD growth of h-BN on Ru(0001). In fact we note that there is some preliminary experimental evidence as shown in \autoref{fig:2DGr}, that an intermediate / precursor structure exists also for CVD of graphene on Ni(111) at temperatures below the best growth conditions. Only upon heating to \SI{730}{\K} the additional diffraction peak vansih, leaving just the $(1\times1)$ graphene/Ni(111) structure behind\cite{tamtogl2015SI}.\vfill

\setlength{\bibsep}{0.0pt}
\putbib[literature]
\end{bibunit}

\end{document}